\documentclass[10pt]{article}
\usepackage{amsmath,amssymb}
\usepackage{graphicx,verbatim,hyperref}
\usepackage[frak=esstix,scr=boondoxo,cal=dutchcal]{mathalfa}
\usepackage{natbib} 
\def\citep{}\renewcommand{\citep}[2][]{(\citealt[#1]{#2})}
\def\citeyearp{}\renewcommand{\citeyearp}[2][]{(\citeyear[#1]{#2})}  
\def\citeyearx{}\renewcommand{\citeyearx}[2][]{(\citeyear[#1]{#2})}  
\def\semi{; }  
\def\paren#1{(#1)}  

\def\eps{\varepsilon}
\def\reps{\check{\varepsilon}}
\def\top#1#2{\genfrac{}{}{0pt}{3}{\raisebox{-1.9pt}{\(\scriptscriptstyle #1\)}}{\raisebox{1.9pt}{\(\scriptscriptstyle #2\)}}}
\def\sop{\top{s}{p}}
\def\sfrac#1#2{\genfrac{}{}{}{1}{#1}{#2}}
\newcommand{\overbar}[1]{\mkern 3.mu\overline{\mkern-3.mu#1\mkern-3.mu}\mkern 3.mu}

\begin{document}

\begin{center}
{\Large\bf Zero-Point Forces in Acoustic Waves}
\vskip5mm
{\bf Laurence J. November}\\
\vskip2mm
La Luz Physics, La Luz NM 88337-0217 USA\\
{\tt{laluzphys@yahoo.com}}\\
May 24, 2019
\end{center}

\begin{abstract}\noindent
By the acousto-optic effect, an acoustic plane wave produces a 1D
index-of-refraction or permittivity wave variation through a medium.  But
adjacent material planes of alternating permittivity should interact due to the
zero-point (ZP) field to produce internal forces, roughly like the Casimir
effect in a stack of regularly spaced discrete conducting plates. The ZP force
in a smoothly varying 1D permittivity wave is modeled and found to consist
mainly of bulk repulsive and double-wavenumber harmonics.  It is stronger than
the Casimir ZP attractive force in the corresponding discrete alternating-layer
stack at all physically meaningful repetition scales, extends to larger scales,
falling off universally only as the inverse square of the wavelength, and shows
no temperature sensitivity.  Thus, at its extremes, a standing acoustic wave
exhibits a bulk expansive ZP pressure through the material volume, but as it
passes through its null the ZP pressure vanishes, giving a body stress modulated
at twice the acoustic wave frequency. But such repeated tensing in a piezo
material is a usual energy-harvesting scenario, suggesting that ZP energy
transfer may occur naturally with standing acoustic waves in a piezo medium.  A
voltage effect is predicted for biphonon lattice vibrations in piezo crystals
with the possibility of `crystal power', the extraction of electrical ZP energy
across the crystal volume.
\vskip2mm
\noindent{\bf keywords:}
Lifshitz theory of Casimir effect; zero-point thermodynamics; quantum
optics:qed; acousto-optical devices.
\end{abstract}

\section{Introduction}
\label{s:intro}

Casimir and Polder \citep{Casimir+Polder1948, Casimir1948} predicted that
discrete closely spaced uncharged parallel conducting plates in a vacuum should
exhibit a force of attraction.  The `Casimir effect' arises due to the
distribution of electromagnetic fluctuations in the background quantum ZP field
conditioned by the flat parallel-plate conductors.

It might seem that with such ZP forces, parallel conducting plates should
require for their separation at least the energy made available in their
attraction. Many physical processes might be said to {\it borrow} intrinsic
energy with the system restored to the original free state with the resupplying
of the same quantity of energy that was abstracted, as in the propelling of a
vehicle against gravity, or in the warming of a liquid in provision of its
latent heat of vaporization for the separating of condensed atoms brought
together by ZP forces. However, unlike gravity or passive ZP effects, the ZP
force between parallel plates can be turned on or off with tunably conductive,
e.g.\ semiconductor plates, so it actually may {\it not} always similarly
symmetrically cycle the ZP energy:

If parallel tunably conductive plates operate elastically as against a
spring-like material between them, the work done in their attraction when the
Casimir force is turned on may be much more than what is needed to drive their
subsequent separation when the Casimir force is turned off.  Energy requirements
for the switching process over a cycle should not be related to the strength of
the ZP compression, as the separation scale is the single most important
property determining the strength of the ZP force.  Thus, in principle, it
should be possible to design an experiment that takes some energy from the ZP
field in each compression cycle.

Various possibilities for the extraction of ZP energy have been put forward
based upon different physical effects \citep{Forward1984, Pinto1999, Pinto2008a,
Maclay2000a, Feigel2004a, Birkeland+Brevik2007, Haisch+Moddel2008}. Cole and
Puthoff \citeyearp{Cole+Puthoff1993} argue that ZP energy transfer for a class
of extraction methods is not in violation of thermodynamic laws.

\begin{figure}[htb!]
\centering\noindent
\centerline{\includegraphics[width=0.6\columnwidth]{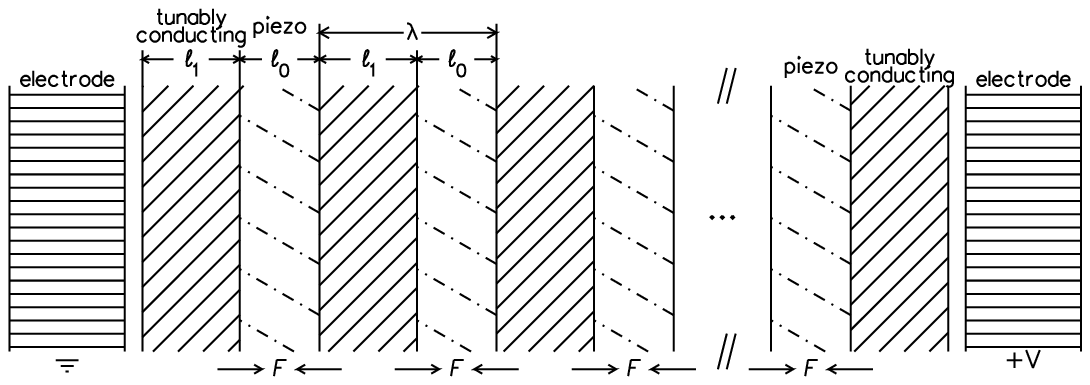}}
\caption{Stack of alternating tunably conductive and piezo plates.  Switching
the tunably conductive plates to conduct and then insulate produces a Casimir ZP
compression of the intermediate piezos followed by their elastic reexpansion,
giving a net work from which useful electrical energy might be derived.}
\label{f:semi}
\end{figure}

To understand the possibilities, a stack consisting of alternating tunably
conductive and piezo layers bounded by conducting electrodes is envisaged, as
illustrated in Figure \ref{f:semi}.  The ZP compression of the piezo produces
both a mechanical spring-like potential, which drives its elastic separation
when the ZP force is turned off, and an electrical potential, which can provide
some energy to an external circuit.  Though models of the Casimir force usually
suppose infinitely thick sandwiching plates, the determining ZP fluctuations are
in the conducting surfaces, with solutions little changed even with plates as
thin as their separations.

The modulation of the Casimir effect by the conductive cycling of semiconductors
with laser light has been demonstrated in precise differential tests, which
validate theoretical estimates to within 1\% \citep{Caride+++2005,
Klimchitskaya+++2009}. Novel semiconductor conductivity switching schemes are
described in the literature \citep{Lipkin1996}. Semiconductor modulation at
rates up to about $\nu=10$ THz by electrical modulation in adjacent oppositely
doped layers for modern transistor design has been demonstrated
\citep{Cooke2007}, and deposited layers only 5 nm thick, about 18 atoms, are
reached in microlayer nanotechnology, like what is used in integrated circuits.
With such rates and scales, the maximum possible ZP power that might be made
available with an ideal perfect conductor in alternating layers is in the range
of 60 kW/cm$^3$ of the stack volume, but drops off very rapidly using realistic
material conductors or semiconductors to ${<}120$ W/cm$^3$, as elaborated in
Appendix \ref{a:stacknum}. Though a possibly viable arrangement, this paper
considers the alternating-layer stack only as a conceptual example to help
understand qualitatively similar, but more natural forms of ZP force modulation
with better possibilities for much more appreciable energy extraction.

\begin{figure}[htb!]
\centering\noindent
\centerline{\includegraphics[width=0.6\columnwidth]{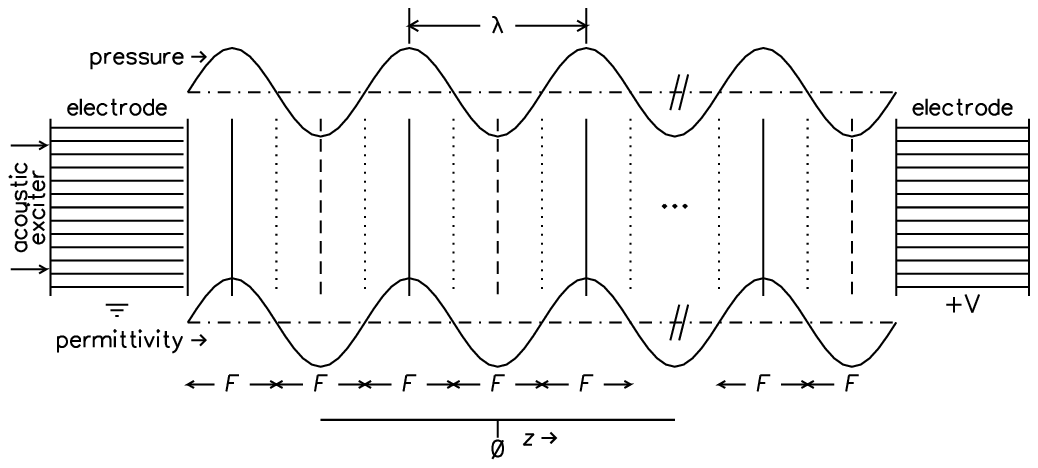}}
\caption{ZP forces in an acoustic wave. Neighboring permittivity plane wave
crests (or troughs) produced in an acoustic wave should interact like
alternating discrete parallel layers to give ZP pressures within the medium.}
\label{f:piezo}
\end{figure}

Here the ZP dynamics of acoustic waves is considered. A plane pressure wave
produces a 1D periodic graded index-of-refraction or permittivity variation by
the acousto-optic effect, giving parallel planes of alternating permittivity
through a medium, as illustrated in Figure \ref{f:piezo}. A permittivity
variation is a generalized form of conductivity variation, so neighboring plane
wave crests (or troughs) should exhibit a ZP force between them, like
alternating discrete conducting plates in a stack, giving longitudinal pressure
variations within the medium, as illustrated by the ZP force vectors $F$.

It appears that ZP forces in acoustic waves have not been discussed before in
the scientific literature, yet it is predicted that effects can be large.  The
`acoustic Casimir effect' is an analogue for the Casimir effect wherein acoustic
waves play the role of the background ZP fluctuating field.  Acoustic waves,
like ZP fluctuations, behave like ocean waves outside nearby ships, which force
them together as their hulls impose destructive interference on the waves
between them, an effect known to ancient mariners for its potentially hazardous
consequences.

ZP forces in graded-permittivity media have been considered mainly to understand
the effects of soft boundaries on three-layer solutions, and unlike discrete
layers, are described by divergent quantum integrals, which may suggest an
enhanced effect \citep{Inui2003a, Podgornik+Parsegian2004a,
Philbin+Xiong+Leonhardt2010}.  The ZP force in a material with a 1D spatially
varying permittivity is formulated in Section \ref{s:cas}, with Fourier-series
solutions derived in Section \ref{s:graded} using an appropriate high-frequency
cutoff to avoid quantum integral divergences, and with numerical examples for a
permittivity wave presented in Section \ref{s:gradnum}.  The ZP force in a
periodic waveform of graded permittivity is indeed found to be greatly enhanced
in amplitude compared to the Casimir ZP force in the otherwise equivalent stack
of discrete layers for spatial repetition scales even as small as atomic
dimensions.  It extends to larger repetition scale $\lambda$ decreasing only as
$1/\lambda^2$, exhibits no temperature sensitivity, and is repulsive in a
sinusoidal wave.

A permittivity wave produces mainly two ZP harmonics:\ of bulk dc and double
wavenumber. As a standing acoustic wave is cycled in time, the ZP force turns on
and off. The same sign of ZP bulk force occurs at both wave extremes, but
vanishes as the standing wave passes through its null, giving a body force
modulated at twice the acoustic wave frequency, as illustrated in the timeline
in Figure \ref{f:timeline}.  But such repeated body tensing in a piezo crystal
is a usual energy-harvesting scenario, suggesting that ZP energy might be
derived from a standing acoustic wave through end electrodes in a `crystal
power' optical arrangement like that illustrated in Figure \ref{f:piezo}. In a
traveling acoustic wave, no ZP energy transfer occurs as the ZP bulk harmonic is
not modulated.

\begin{figure}[htb!]
\centering\noindent
\centerline{\includegraphics[width=0.5\columnwidth]{timeline.eps}}
\caption{Cycling of ZP pressure in a standing acoustic wave of frequency $\nu$
cycles per sec in time $t$.  The ZP bulk repulsive (expansive) force in a
standing permittivity wave is modulated at twice the acoustic wave frequency.}
\label{f:timeline}
\end{figure}

The ZP bulk repulsive force in a standing acoustic wave may be represented as a
traction force acting within the volume of the piezo crystal, which produces
both stress and strain mechanical variations as discussed in Section
\ref{s:piezo}.  The piezo effect introduces a shift in the temporal phase
between the mechanical stress and strain cycles and correspondingly between the
electric-field and electric-displacement cycles, representing a transfer of
energy in a conservative system.  The energetic coupling of the mechanical cycle
to an external electrical load (or supply), shifts the stress cycle in phase to
lead (or lag) the strain, producing a single-signed cycling electric
displacement that leads (or lags) a single-signed oscillating electric field,
which represents a systematic transfer of energy out of (or into) the mechanical
cycle, driven by (or in opposition to) ZP forces.

As the ZP bulk pressure produces a single-signed piezo voltage across the
crystal volume, even the out-of-phase modulation in finite domains should add to
give a net voltage effect.  Possibilities for the excitation of standing
acoustic waves in piezo crystals relevant for a voltage effect are discussed in
Section \ref{s:disc}.

\section{1D ZP Force}
\label{s:cas}

ZP forces arise in a medium with a 1D spatial $z$ permittivity variation
$\eps[\xi,z]$ as in an alternating-layer stack like that illustrated in Figure
\ref{f:semi}, or in an acoustic wave in Figures \ref{f:piezo} and
\ref{f:timeline}.  The general complex permittivity $\eps[\xi,z]$ characterizes
fully the electro-optic properties of a medium, defining its retardation and
absorption as a function of light frequency $\xi$.  The ZP force is derived in
many places especially after the Lifshitz collaboration (\citealt{Lifshitz1956,
Dzyaloshinskii+++1961}; see \citealt{Milonni1994}, Chapter 7), and measurements
like those described in the introduction, validate their solution in retarding
and absorbing media.  The ZP force per unit area (or pressure) with positive as
expansive is written as a function of position $z$
\begin{equation}
F[z]=-\frac{k_{\rm{B}}\Theta}{\pi c^2}\sum_{m=0}^\infty{\left(\sfrac{1}{2}\right)^{(m==0)}
\int_0^\infty{\kappa[z]\left(\mathcal{R}_s[z]+\mathcal{R}_p[z]\right)\omega d\omega}},
\label{e:lif}
\end{equation}
for $k_{\rm{B}}$ Boltzmann's constant and $\Theta$ the temperature.  The sum
over $m$ counts the Matsubara mode frequencies $\xi=m 2\pi
k_{\rm{B}}\Theta/\hslash=m(\Theta$/300K)$\cdot2.47$E14 rad/s, which enter
implicitly under the integral; as usual $h=2\pi\hslash$ is Planck's constant and
$c$ the speed of light.  The logical function is used with $(m{==}0)=1$ on true
or 0 on false, giving an $m=0$ summation term that receives half weight. For
ease in reading the complicated relations, square braces are used throughout
this paper to denote functional dependencies.

The reflection terms for perpendicular (senkrecht) $s$ and parallel $p$ ZP
polarized rays are defined
\begin{equation}
\mathcal{R}_{\sop}[z]=
\frac{R_{\sop+}[z]R_{\sop-}[z]}{1-R_{\sop+}[z]R_{\sop-}[z]},
\label{e:RRsp}
\end{equation}
for Fresnel reflection coefficients $R_{\sop\pm}[z]$, which depend upon the
spatial position $z$ as well as the permittivity.  The wavenumber $\kappa[z]$
for the ZP evanescent electromagnetic modes is defined by the mode frequencies
$\xi$ and $\omega$
\begin{equation}
\kappa[z]=\frac{1}{c}\left(\reps[\xi,z]\xi^2+\omega^2\right)^\frac{1}{2}.
\label{e:kap}
\end{equation}
The definitions for the real wavenumber $\kappa[z]$ and Fresnel reflection
coefficients $R_{\sop\pm}[z]$ are based upon the real permittivity
$\reps[\xi,z]\equiv\eps[i\xi,z]$, the projection of the general complex
permittivity on its imaginary frequency axis, which casts the complex
$\eps[i\xi]$ into a real function $\reps[\xi]$ of real frequency $\xi$
consistent with the Kramers-Kr{\"o}nig causality constraints
(\citealt{Hough+White1980}; see Appendix \ref{a:stacknum}).  Fresnel reflection
coefficients for evanescent waves defined using the real projected permittivity
$\reps_{\jmath}[\xi]$ give a convenient real solution form for mixed retarding
and absorbing multilayers.  Variables retain dependencies on frequencies
implicitly as a dependence on $\xi$ in the permittivity $\reps[z]$, and
dependencies on $\xi$ and $\omega$ in the wavenumber $\kappa[z]$.

It is usual to replace the sum over Matsubara frequencies in (\ref{e:lif}) by an
integral. Symbolically the sum is approximated
$\sum_{m=0}^\infty{(1/2)^{(m==0)}\cdots}\rightarrow\int_0^\infty{\cdots dm}$,
which takes properly into account the half-size interval at the lower $m=0$
limit.  Substituting for the sum in this way, and using the interval
$dm=(\hslash/(2\pi k_{\rm{B}}\Theta))d\xi$ from the definition for the Matsubara
frequency $\xi$, gives the integral for the ZP force
\begin{equation}
F[z]=-\frac{\hslash}{2\pi^2c^2}\int_{0}^\infty{
\int_0^\infty{\kappa[z]\left(\mathcal{R}_s[z]+\mathcal{R}_p[z]\right)\omega d\omega}d\xi},
\label{e:lifx01}
\end{equation}
which is independent of temperature $\Theta$, and so represents the
zero-temperature $\Theta=0$K solution.

Simply substituting an integral for the sum is a good approximation when the
main contribution to the sum comes from terms with $m\gg1$.  Thus temperature
effects are only important when terms of low frequency $\xi$ or small wavenumber
$\kappa$ are significant in sum, which occurs when the layer thickness $\ell$ is
comparable to or greater than the light wavelength for the Matsubara base
frequency $\ell\gtrsim c/\xi[m{=}1]=c/(2\pi
k_{\rm{B}}\Theta/\hslash)=(300\rm{K}/\Theta)\cdot1.2148$E${-}4$ cm.  As is
illustrated in numerical examples in Figure \ref{f:ftemp} in Appendix
\ref{a:stacknum}, temperature effects in multilayer solutions are indeed of
vanishing importance when the middle-layer thickness is less than about E${-}4$
cm = 1 $\mu$m.  The summation for finite temperature effects can be considered
via Abel-Plana formulae \citep{Dowling1989}.

\begin{figure}[htb!]
\centering\noindent
\centerline{\includegraphics[width=0.5\columnwidth]{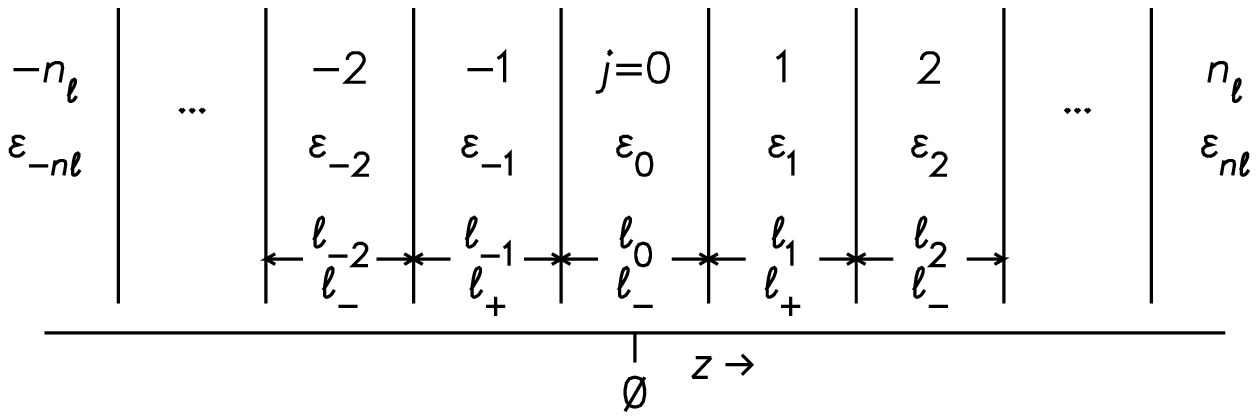}}
\caption{Plane-parallel $2n_{\ell}+1$ multilayer stack consisting of projected
permittivities $\reps_{\jmath}$ and wavenumbers $\kappa_{\jmath}$ in layers
$\jmath$ counted from a middle layer $\jmath=0$, with interfacial reflections
$r_{s\jmath}$ and $r_{p\jmath}$ at the jumps $z_{\jmath}$ between layers.}
\label{f:stack}
\end{figure}

The Lifshitz collaborative QED formulation given in Dzyaloshinskii et al.\
\citeyearp{Dzyaloshinskii+++1961} has been shown to be applicable in a 1D stack
with an arbitrary number of multilayers \citep{Ninham+Parsegian1970a,
Podgornik+Hansen+Parsegian2003}. Fresnel reflection coefficients are propagated
in $+z$ by recursion as $R_{\sop+}[z]$ from $z=-\infty$ or in $-z$ as
$R_{\sop-}[z]$ from $z=+\infty$ following the comprehensive formula for
evanescent waves
\paren{\citealt{Jacobsson1965}\semi\citealt{Zhou+Spruch1995}\semi\citealt[Section
1.6]{Born+Wolf1980}}.  The $+z$ propagated reflection coefficient changes from
$R_{\sop+}[{\prec}z_{\jmath}]$ just below the permittivity jump at $z_{\jmath}$
up to the next following jump at $z_{\jmath+1}$ as
\begin{equation}
R_{\sop+}[z_{\jmath}{<}z{<}z_{\jmath+1}]=\frac{R_{\sop+}[{\prec}z_{\jmath}]+r_{\sop\jmath}}
{1+R_{\sop+}[{\prec}z_{\jmath}]r_{\sop\jmath}}e^{-2\kappa_{\jmath}(z-z_{\jmath})},
\label{e:Rsp+}
\end{equation}
using $R_{\sop+}[{\prec}z_{\jmath}]$ to denote the value of $R_{\sop+}$
preceding or smoothly asymptotic with $z=z_{\jmath}$ from below.  The
interfacial reflections are defined
\begin{equation}
r_{s\jmath}=\frac{\kappa_{\jmath-1}-\kappa_{\jmath}}{\kappa_{\jmath-1}+\kappa_{\jmath}},\quad\quad
r_{p\jmath}=\frac{\reps_{\jmath}\kappa_{\jmath-1}-\reps_{\jmath-1}\kappa_{\jmath}}
{\reps_{\jmath}\kappa_{\jmath-1}+\reps_{\jmath-1}\kappa_{\jmath}},
\label{e:rsp}
\end{equation}
for permittivities $\reps_{\jmath}$ and wavenumbers $\kappa_{\jmath}$ in layers
numbered $\jmath$, with $\ell_{\jmath}$ the layer thickness as illustrated in
Figure \ref{f:stack}.  In a stack with equal layer thicknesses
$\ell_{\jmath}=\ell$, the jumps are located at
$z_\jmath=(\jmath-\frac{1}{2})\ell$, $z_{-1}=-\sfrac{3\ell}{2}$,
$z_0=-\sfrac{\ell}{2}$, $z_1=\sfrac{\ell}{2}$, $z_2=\sfrac{3\ell}{2}$, etc.
Alternating-layer arrangements are considered too, which have two thicknesses
with $\ell_{\jmath}=\ell_0$ in even $\jmath$ layers and $\ell_{\jmath}=\ell_1$
in odd $\jmath$ layers.

Similarly the $-z$ propagated reflection coefficient changes from
$R_{\sop-}[{\succ}z_{\jmath+1}]$ succeeding or just above the boundary between
layers $\jmath+1$ to $\jmath$, to smaller $z$ down to the next boundary at
$z_{\jmath}$, with the formula
\begin{equation}
R_{\sop-}[z_{\jmath+1}{>}z{>}z_{\jmath}]=\frac{R_{\sop-}[{\succ}z_{\jmath+1}]-r_{\sop(\jmath+1)}}
{1-R_{\sop-}[{\succ}z_{\jmath+1}]r_{\sop(\jmath+1)}}e^{-2\kappa_{\jmath}(z_{\jmath+1}-z)},
\label{e:Rsp-}
\end{equation}
with the interfacial reflections across the jump defined in (\ref{e:rsp}),
noting reversed signs for the reversed propagation direction in $z$ compared to
Eq.\ (\ref{e:Rsp+}).

The Fresnel reflection coefficients decrease exponentially in $+z$ from
$R_{\sop+}[{\succ}\sfrac{-\ell_0}{2}]=(R_{\sop+}[{\prec}\sfrac{-\ell_0}{2}]+r_{\sop0})/
(1+R_{\sop+}[{\prec}\sfrac{-\ell}{2}]r_{\sop0})$ just above the
$z_0=\sfrac{-\ell_0}{2}$ jump in Eq.\ (\ref{e:Rsp+}), and in $-z$ from
$R_{\sop-}[{\prec}\sfrac{\ell_0}{2}]=(R_{\sop-}[{\succ}\sfrac{\ell_0}{2}]-r_{\sop1})/
(1-R_{\sop-}[{\succ}\sfrac{\ell}{2}]r_{\sop1})$ just below the
$z_1=\sfrac{\ell_0}{2}$ jump in Eq.\ (\ref{e:Rsp-}).  They vary across the middle
layer $\sfrac{-\ell_0}{2}<z<\sfrac{\ell_0}{2}$ as
\begin{equation}
R_{\sop+}[z]=R_{\sop+}[{\succ}\sfrac{-\ell_0}{2}]e^{-2\kappa_0\left(z+\frac{\ell_0}{2}\right)},\quad\quad
R_{\sop-}[z]=R_{\sop-}[{\prec}\sfrac{\ell_0}{2}]e^{-2\kappa_0\left(\frac{\ell_0}{2}-z\right)},
\nonumber
\end{equation}
and in their product, the $z$ dependence vanishes
\begin{equation}
R_{\sop+}[z]R_{\sop-}[z]=
R_{\sop+}[{\succ}\sfrac{-\ell_0}{2}]R_{\sop-}[{\prec}\sfrac{\ell_0}{2}]e^{-2\kappa_0\ell_0},
\label{e:RRz0}
\end{equation}
giving $\mathcal{R}_{\sop}$ from (\ref{e:RRsp}) and a Casimir force $F$ from
(\ref{e:lif}) or (\ref{e:lifx01}) independent of $z$ across a constant
permittivity layer within jumps on either side.  A constant ZP force across
a uniform layer is a well-known property of ZP force solutions.

A three-layer stack has the Fresnel reflection coefficient
$R_{\sop+}[{\succ}\sfrac{-\ell_0}{2}]=r_{\sop0}$ above its lower jump at
$z=\sfrac{-\ell_0}{2}$, projecting with Eq.\ (\ref{e:Rsp+}) from a uniform media
everywhere below $R_{\sop+}[z{<}\sfrac{-\ell_0}{2}]=0$.  It has the reflection
coefficient $R_{\sop-}[{\prec}\sfrac{\ell_0}{2}]=-r_{\sop1}$ below its upper jump
at $z=\sfrac{\ell_0}{2}$, projecting with (\ref{e:Rsp-}) from a uniform media
everywhere above $R_{\sop-}[z{>}\sfrac{\ell_0}{2}]=0$. Combining gives the product
$R_{\sop+}[z]R_{\sop-}[z]=-r_{\sop0}r_{\sop1}\exp[-2\kappa_0\ell_0]$ in the middle
layer $\sfrac{-\ell_0}{2}<z<\sfrac{\ell_0}{2}$ from (\ref{e:RRz0}).  With
(\ref{e:rsp}), the product leads to the Lifshitz collaborative formula
\citep[Eq.\ (4.14)]{Dzyaloshinskii+++1961}.

A five-layer stack $n_{\ell}=2$ with layers of equal thicknesses $\ell$ counted
$-2\le j\le2$ has four contained jumps, at $z_{-1}=\sfrac{-3\ell}{2}$,
$z_0=\sfrac{-\ell}{2}$, $z_1=\sfrac{\ell}{2}$, and $z_2=\sfrac{3\ell}{2}$ as in
Figure \ref{f:stack}.  The Fresnel reflection coefficient in (\ref{e:RRz0})
comes from the $+z$ propagated coefficient just above the jump at $z_{-1}$,
$R_{\sop+}[{\succ}\sfrac{-3\ell}{2}]=r_{\sop-1}$, which gives
$R_{\sop+}[{\prec}\sfrac{-\ell}{2}]=r_{\sop-1}\exp[-2\kappa_{-1}\ell]$ just
below the next jump at $z_0$, and then the needed
$R_{\sop+}[{\succ}\sfrac{-\ell}{2}]$ across $z_0$ with $r_{\sop0}$ using Eq.\
(\ref{e:Rsp+}).  Similarly the Fresnel reflection coefficient just below the
jump at $z_2$, $R_{\sop-}[{\prec}\sfrac{3\ell}{2}]=-r_{\sop2}$, is propagated in
$-z$ to yield
$R_{\sop-}[{\succ}\sfrac{\ell}{2}]=-r_{\sop2}\exp[-2\kappa_{1}\ell]$ just above
the next lower jump at $z_1$, and then the needed
$R_{\sop-}[{\prec}\sfrac{\ell}{2}]$ below $z_1$ with $r_{\sop1}$ using Eq.\
(\ref{e:Rsp-}).  The resulting product of reflection coefficients in Eq.\
(\ref{e:RRz0}) agrees with Zhou and Spruch \citeyearx[Eqs.\
(3.12)--(3.16)]{Zhou+Spruch1995}.

Though changing symbols and formulae add serious complications, the Casimir
force in (\ref{e:lif}) or (\ref{e:lifx01}) with any number of layers appears to
be consistent with what is developed and discussed in numerous studies
\citep{Tomas2002, Raabe+++2003a, Henkel+Joulain2005, Ellingsen2007}.

A sandwich of two identical layers around a middle layer $\jmath=0$ with
permittivities $\reps_{1}=\reps_{-1}$ has jump coefficients
$r_{\sop0}=-r_{\sop1}$, and so a product of reflection coefficients
$R_{\sop+}[z]R_{\sop-}[z]=r_{\sop0}^2\exp[-2\kappa_0\ell_0]>0$ or
$\mathcal{R}_{\sop}[z]>0$ from (\ref{e:RRsp}), meaning that the Casimir ZP force
$F[z]$ defined in Eqs. (\ref{e:lif}) or (\ref{e:lifx01}) is negative or
attractive.  It can be seen too that recursive application of the formulae
(\ref{e:Rsp+}) and (\ref{e:Rsp-}) in a spatially symmetric arrangement around a
middle layer $\jmath=0$ with $\reps_{\jmath}=\reps_{-\jmath}$, for any number of
layers even of varying thicknesses with $\ell_{\jmath}=\ell_{-\jmath}$, gives
$R_{\sop+}[{\succ}\sfrac{-\ell_0}{2}]=R_{\sop-}[{\prec}\sfrac{\ell_0}{2}]$, and
so in the middle layer $R_{\sop+}[z]R_{\sop-}[z]>0$, $\mathcal{R}_{\sop}[z]>0$,
and an attractive ZP force $F[z]<0$ (as in
\citealt{Lambrecht+Jaekel+Reynaud1997, Henkel+Joulain2005}). For a symmetric
arrangement it also follows that $R_{\sop+}[z]=R_{\sop-}[-z]$ or
$R_{\sop+}[z]R_{\sop-}[z]=R_{\sop+}[-z] R_{\sop-}[-z]$, giving
$\mathcal{R}_{\sop}[z]=\mathcal{R}_{\sop}[-z]$, and so a symmetric ZP force
$F[z]=F[-z]$ for all $z$.

In principle the $\omega$ integral in Eqs.\ (\ref{e:lif}) and (\ref{e:lifx01})
may be divergent, as the interfacial reflections $r_{p\jmath}$ from
(\ref{e:rsp}) do not vanish as $\omega\rightarrow\infty$, and these enter into
the reflection coefficients $R_{p\pm}[z]$ and reflection term $\mathcal{R}_{p}$.
However the attenuation factor $\exp[-2\kappa_0\ell_0]$, which multiplies the
product of the reflection coefficients from the two nearest jumps in
(\ref{e:RRz0}), does go to zero as $\omega$ goes to infinity for finite
$\ell_0$, since $\kappa_0$ from Eq.\ (\ref{e:kap}) increases with $\omega$.
Thus quantum divergences are not seen where the permittivity is constant across
a finite layer.

For a three-layer stack, the ZP force integral Eq.\ (\ref{e:lifx01}) in a middle
layer of thickness $\ell_0$ is conveniently rewritten in terms of a new
dimensionless integration variable
$\eta=\kappa_0\ell_0=(\reps[\xi]\xi^2+\omega^2)^{1/2}\ell_0/c$, which
substitutes for the frequency $\omega$.  Using the derivative
$\partial\eta^2/\partial\omega^2=(\ell_0/c)^2$ or $\eta d\eta=(\ell_0/c)^2\omega
d\omega$, reforms the $\omega$ integral into an integral in $\eta$ as
\begin{equation}
F=-\frac{\hslash}{2\pi^2\ell_0^3}\int_{0}^\infty{
\int_0^\infty{\eta^2\left(\mathcal{R}_s+\mathcal{R}_p\right)d\eta}d\xi}.
\label{e:lifx02}
\end{equation}
As described in Appendix \ref{a:limits}, for a small middle-layer thickness
$\ell_0$, the $\eta$ integrand is determined by the product
$\eta^2\exp[-2\eta]$, and the ZP force variation with $\ell_0$ follows the
external factor $1/\ell_0^3$. The product $\eta^2\exp[-2\eta]$ has its maximum
around $\eta=1$, corresponding to the wavenumber $\kappa_0=1/\ell_0$, and drops
off exponentially beyond.

Other limiting forms arise when the middle-layer thickness is greater than a
cross-over scale $\ell_0>\ell_{\pm}$, as most usually a $1/\ell_0^4$ ZP force
retarded dependence as elaborated in Appendices \ref{a:stacknum} and
\ref{a:limits}. Symmetric multilayer stacks exhibit the same asymptotic behavior
as a three-layer stack, since ZP forces are determined in the nearest layers.

\section{In a Graded Permittivity}
\label{s:graded}

The definition for the ZP force from Eq.\ (\ref{e:lif}) should be applicable too
in a stratified medium with a 1D smoothly varying complex permittivity $\eps[z]$
or real $\reps[z]$ and corresponding real $\kappa[z]$ defined by Eq.\
(\ref{e:kap}).  In the limit that the layer thickness $\ell$ becomes
infinitesimal, the Fresnel recursion formulae Eqs.\ (\ref{e:Rsp+}) and
(\ref{e:Rsp-}) produce differential equations for a graded permittivity.

The expression for the reflection coefficient at the high $z$ end of the
$\jmath=0$ interval $R_{\sop\pm}[{\prec}\sfrac{\ell}{2}]$ in terms of its value
just below the low $z$ jump $R_{\sop\pm}[{\prec}\sfrac{-\ell}{2}]$ from Eq.\
(\ref{e:Rsp+}), with the substitution of Taylor series' in powers of $\ell$,
$R_{\sop+}[{\prec}\pm\sfrac{\ell}{2}]=R_{\sop+}[0]\pm({\partial
R_{\sop+}/\partial z})\sfrac{\ell}{2}+\cdots$,
$r_{\sop0}=\rho_{\sop}[-\sfrac{\ell}{2}]\ell=\rho_{\sop}[0]\ell-({\partial\rho_{\sop}/\partial
z})\ell^2/2+\cdots$, and $\exp[-2\kappa_0\ell]=1-2\kappa_0\ell+\cdots$, gives
the differential equation for $R_{\sop+}[z]$ around $z=0$
\begin{equation}
\frac{\partial R_{\sop\pm}}{\partial z}=\rho_{\sop}[z](1-R_{\sop\pm}[z]^2)\mp 2\kappa[z] R_{\sop\pm}[z],
\label{e:dRsp}
\end{equation}
introducing the smoothly varying reflection $\rho_{\sop}[z_{\jmath}]\equiv
r_{\sop\jmath}/\ell$, and keeping terms only to lowest order in $\ell$.  The
differential equation for $R_{\sop-}[z]$ is contained in this equation also,
which is found similarly by defining the reflection coefficient for the low $z$
end of the $\jmath=0$ interval $R_{\sop-}[{\succ}\sfrac{-\ell}{2}]$ in terms of
its value just above the high $z$ jump $R_{\sop-}[{\succ}\sfrac{\ell}{2}]$ using
Eq.\ (\ref{e:Rsp-}), with the additional Taylor series'
$R_{\sop-}[{\succ}\pm\sfrac{\ell}{2}]=R_{\sop-}[0]\pm({\partial
R_{\sop-}/\partial z})\sfrac{\ell}{2}+\cdots$, and
$r_{\sop1}=\rho_{\sop}[\sfrac{\ell}{2}]\ell=\rho_{\sop}[0]\ell+({\partial\rho_{\sop}/\partial
z})\ell^2/2+\cdots$.  Though relations are derived for around $z=0$, they are
applicable at all $z$ in the small $\ell$ limit for an infinite multilayer
stack.

The smoothly varying reflection $\rho_{\sop}[z_{\jmath}]\equiv
r_{\sop\jmath}/\ell$ is introduced as it remains well defined in the small
$\ell$ limit.  The wavenumber and permittivity first derivatives are taken to be
the jump differences between layers normalized by the layer thickness $\ell$, as
$\partial\kappa/\partial z=(\kappa_0-\kappa_{-1})/\ell$ and
$\partial\reps/\partial z=(\reps_0-\reps_{-1})/\ell$. Substituting into Eq.\
(\ref{e:rsp}) with the values at $z=0$, $\kappa[0]=\kappa_0$ and
$\reps[0]=\reps_0$, then yields around $z=0$
\begin{equation}\begin{split}
\rho_s[z]&=-\frac{1}{2\kappa[z]}\frac{\partial\kappa}{\partial z}=
-\left(\frac{\xi}{2c\kappa[z]}\right)^2\frac{\partial\reps}{\partial z},\\
\rho_p[z]&=\frac{1}{2\reps[z]}\frac{\partial\reps}{\partial z}-
\frac{1}{2\kappa[z]}\frac{\partial\kappa}{\partial z}
=\left(\frac{1}{2\reps[z]}-\left(\frac{\xi}{2c\kappa[z]}\right)^2\right)
\frac{\partial\reps}{\partial z},
\label{e:rhosp}
\end{split}\end{equation}
written for $\ell\rightarrow0$, and again recognizing that the relations remain
applicable at all $z$ in an infinite multilayer stack.  The differential Eq.\
(\ref{e:dRsp}) with (\ref{e:rhosp}) agrees with evanescent solutions from
different formulations for the propagation of waves in a stratified medium
(e.g.\ see \citealt[Eq.\ (2.37)]{Jacobsson1965}\semi \citealt[Chapter
5]{Lekner2016}\semi and references).

For everywhere-small reflection coefficients $|R_{\sop\pm}[z]|\ll1$, consistent
with small-amplitude periodic permittivity variations on top of a constant
permittivity background, Eq.\ (\ref{e:dRsp}) can be linearized, which provides
the closed-form integral solution
\begin{equation}
R_{\sop\pm}[z]=e^{\mp2\int{\kappa[z]dz}}
\int{e^{\pm2\int{\kappa[z]dz}}\rho_{\sop}[z]
\left(1-\tilde{R}_{\sop\pm}[z]^2\right)dz},
\label{e:Rint}
\end{equation}
where the function $\tilde{R}_{\sop\pm}[z]$ represents a prior approximation to
the propagated Fresnel reflection coefficient.

ZP forces in a medium of graded permittivity are studied by Fourier analysis
supposing relatively small reflection coefficients, which gives a denominator
near one in $\mathcal{R}_{\sop}[z]$ from (\ref{e:RRsp}), and a ZP force from
(\ref{e:lifx01}) well approximated as
\begin{equation}
F[z]=-\frac{\hslash}{2\pi^2c^2}\int_0^\infty{\int_0^\infty{\kappa[z]\left(R_{s+}[z]R_{s-}[z]+
R_{p+}[z]R_{p-}[z]\right)\omega d\omega}d\xi}.
\label{e:lifx03}
\end{equation}
The approximations used here are quantified in numerical calculations described
in Section \ref{s:gradnum}.

The Fourier transform of a repeating spatial profile of some repetition length
$\lambda$ can be expanded in a discrete Fourier series in wavenumbers
$n/\lambda$ cycles per cm for integer modes $n$, extending in principle from
$-\infty<n<\infty$, written for the ZP force as
\begin{equation}
F[z]=\sum_n{\bar{F}_n e^{i2\pi n\frac{z}{\lambda}}},
\label{e:Fser}
\end{equation}
in Fourier modes $\bar{F}_n$, or similarly for the reflection coefficients
$R_{\sop\pm}[z]$ in modes $\bar{R}_{(\sop\pm)n}$, or for the smoothly varying
reflections $\rho_{\sop}[z]$ in modes $\bar{\rho}_{\sop n}$.  The Fourier modes
are Hermitian with $\bar{F}_{-n}=\bar{F}_n^*$ for a real function $F[z]$, or are
real with $\bar{F}_{-n}=\bar{F}_n$ for a symmetric function around $z=0$,
$F[z]=F[-z]$.

Substituting Fourier series' for the variables transforms the differential Eq.\
(\ref{e:dRsp}) into the algebraic equation in Fourier modes
\begin{equation}
\bar{R}_{(\sop\pm)n}=\frac{\bar{\rho}_{\sop n}}{\pm2\kappa+i2\pi n/\lambda},
\label{e:FR}
\end{equation}
ignoring the square of the reflection coefficient, and spatial variations in the
wavenumber $\kappa$.  Solutions for the ZP force are found with $\kappa$ in Eq.\
(\ref{e:kap}) determined by a diverging frequency $\omega$, which is spatially
constant.

With $\kappa$ greater than the largest significant wavenumbers $n/\lambda$ in
the smoothly varying reflection, Eq.\ (\ref{e:FR}) gives the proportionality
$\bar{R}_{(\sop\pm)n}\sim\bar{\rho}_{\sop n}/\kappa$ in every important Fourier
mode $n$, which means spatial quantities similarly go as
$R_{\sop\pm}\sim\rho_{\sop}/\kappa$, and the product of reflection coefficients
as $R_{\sop+}R_{\sop-}\sim\rho_{\sop}^2/\kappa^2$.  From Eq.\ (\ref{e:rhosp}),
$\rho_s\sim1/\kappa^2$, and the product of reflection coefficients goes as
$R_{s+}R_{s-}\sim1/\kappa^6$.  On the other hand, the first term in the smoothly
varying reflection $\rho_p$ from Eq.\ (\ref{e:rhosp}) is independent of
$\kappa$, which gives $R_{p+}R_{p-}\sim1/\kappa^2$ asymptotically for large
$\kappa$. All in all, the first term in $\rho_p$ leads to a linearly divergent
$\omega$ integral in Eq.\ (\ref{e:lifx03}), whereas integrals of the other
cross-product terms and of the product $R_{s+}R_{s-}$ are convergent, and can be
ignored by comparison. Thus only the first term in the parallel smoothly varying
reflection determines the ZP force
\begin{equation}
\rho_p[z]=\frac{1}{2}\frac{\partial\ln[\reps]}{\partial z}.
\label{e:rhop}
\end{equation}

So, with a smoothly varying permittivity, the integral over $\omega$ in
(\ref{e:lifx03}) is indeed divergent and feels its maximum contribution from
frequencies $\omega$ approaching a cutoff $\omega_{\rm{x}}$.  On the other hand,
the Matsubara frequency $\xi$ has a limited range of significant contribution as
the spatial derivative of the permittivity $\reps$, and so the smoothly varying
reflection $\rho_p[z]$ in Eq.\ (\ref{e:rhop}), disappear well below any
physically reasonable cutoff with $\xi\ll\omega_{\rm{x}}$, and with them the
reflection coefficient $R_{p\pm}$ from the Fourier relation Eq.\ (\ref{e:FR}).
Thus the wavenumber in (\ref{e:kap}) is well approximated as $\kappa=\omega/c$.
This result is quite unlike what was found with a constant permittivity in a
finite layer $\ell$ where the ZP force integral (\ref{e:lifx02}) has its maximum
contribution around the relatively small wavenumber $\kappa=1/\ell$.

Expressing the permittivity as a Fourier series' too, but in its logarithm
\begin{equation}
\ln[\reps[z]]=\sum_n{\overbar{\{\ln\reps\}}_n e^{i2\pi n\frac{z}{\lambda}}},
\label{e:lepsser}
\end{equation}
for complex coefficients $\overbar{\{\ln\reps\}}_n$, yields for the smoothly
varying reflection from Eq.\ (\ref{e:rhop}), and for the reflection coefficient
from (\ref{e:FR})
\begin{equation}
\bar{\rho}_{pn}=\frac{i \pi n}{\lambda}\overbar{\{\ln\reps\}}_n,\quad\quad
\bar{R}_{(p\pm)n}=\frac{1}{2}\ \frac{in}{\pm\frac{\omega\lambda}{\pi c}+in} \overbar{\{\ln\reps\}}_n,
\label{e:rhoRpm}
\end{equation}
using $\kappa=\omega/c$. Substituting into the ZP force integral Eq.\
(\ref{e:lifx03}) with $R_{s\pm}=0$ and the Fourier series' for the reflection
coefficient $R_{p\pm}[z]$ in its modes $\bar{R}_{(p\pm)n}$ from Eq.\
(\ref{e:rhoRpm}), then gives a Fourier series for the ZP pressure Eq.\
(\ref{e:Fser}) with the mode coefficients
\begin{equation}
\bar{F}_n=\frac{\hslash}{8\pi^2c^3}\sum_q{
\int_0^\infty{(n-q)q\ \overbar{\{\ln\reps\}}_{n-q}\overbar{\{\ln\reps\}}_q\ d\xi}
\int_0^\infty{\frac{\omega^2}{\biggl(\frac{\omega\lambda}{\pi c}+in-iq\biggr)\biggl(\frac{\omega\lambda}{\pi c}+iq\biggr)
\left(1+\left(\frac{\omega}{\omega_{\rm{x}}}\right)^2\right)^2}d\omega}}.
\label{e:lifx04}
\end{equation}
where the sum over $\pm$ integers $q$ represents a convolution, which arises
with the product of Fourier series'. The $\omega$ integral has a smooth
Lorentzian cutoff around an upper limit $\omega_{\rm{x}}$ introduced with the
inserted term in divisor $(1+(\omega/\omega_{\rm{x}})^2)^2$, which imposes a
form of covariant regularization \paren{\citealt[Section 9.6, after Eq.\
(9-89)]{Feynman+Hibbs1965}\semi\citealt[Section 9.2, Eq.\
(9.9)]{Mandl+Shaw2010}}.  The light wavelength for the cutoff is correspondingly
$\lambda_{\rm{x}}=2\pi c/\omega_{\rm{x}}$.

All explicit dependencies upon the repetition scale $\lambda$ factor out of the
$\omega$ integral, substituting for $\omega$ with a dimensionless
$\hat{\omega}=\omega\lambda_{\rm{x}}/(\pi c)$, and for $n$ and $q$ with the
scaled-down $\hat{n}=n\lambda_{\rm{x}}/\lambda$ and
$\hat{q}=q\lambda_{\rm{x}}/\lambda$. The frequency ratio has the equivalences
$\omega/\omega_{\rm{x}}=\lambda_{\rm{x}}/\lambda=\hat{\omega}/\hat{\omega}_{\rm{x}}$,
where $\hat{\omega}_{\rm{x}}=\omega_{\rm{x}}\lambda_{\rm{x}}/(\pi c)=2$.  With
some rearranging, the ZP force mode becomes
\begin{equation}
\bar{F}_n=-\frac{\pi^2\hslash}{16\lambda_{\rm{x}}\lambda^2}\sum_q{(n-q)q\ \mathfrak{C}
\int_0^\infty{\overbar{\{\ln\reps\}}_{n-q}\overbar{\{\ln\reps\}}_q d\xi}},
\label{e:lifx05}
\end{equation}
with
\begin{equation}
\mathfrak{C}=\frac{32}{\pi}\int_{0}^\infty{\frac{\hat{\omega}^2}{(\hat{\omega}-i\hat{n}+i\hat{q})
(\hat{\omega}+i\hat{q})(\hat{\omega}-2i)^2(\hat{\omega}+2i)^2}d\hat{\omega}}.
\label{e:lifx05c}
\end{equation}

The Fourier modes of the physical (real) quantities are Hermitian,
$\bar{F}_{-n}=\bar{F}_{n}^*$ and
$\overbar{\{\ln\reps\}}_{-n}=\overbar{\{\ln\reps\}}_{n}^*$, but the cutoff
function $\mathfrak{C}$ is real, and does not change by taking the complex
conjugate of Eq.\ (\ref{e:lifx05}), and so must remain unchanged with the
equivalent flipping of the signs of both $n$ and $q$.  It follows that the
$\hat{\omega}$ frequency integrand in $\mathfrak{C}$ must also be unchanged with
the change of sign $\hat{\omega}\rightarrow-\hat{\omega}$, which gives for the
integral $\int_{-\infty}^0\cdots d\hat{\omega}=\int_0^{\infty}\cdots
d\hat{\omega}$ as it appears in Eq.\ (\ref{e:lifx05c}). The integral in the
cutoff function is doubled as its domain is extended to span the full
$\pm\hat{\omega}$ real axis. Looping over the upper (or equivalently lower) half
complex $\hat{\omega}$ plane at infinity, which incurs no additional
contribution, provides a contour in the continued $\hat{\omega}$ complex plane
that encloses the $\hat{\omega}=2i$ Cauchy residue plus contained or overlapping
residues for $\hat{n}-\hat{q}>0$ and $\hat{q}<0$. A real resultant for
$\mathfrak{C}$ is found as shown in Figure \ref{f:int}, which ranges from 1 at
the dc $\hat{n}-\hat{q}=\hat{q}=0$, diminishes for larger wavenumbers, and
becomes slightly negative for $(\hat{n}-\hat{q})\hat{q}\lesssim-2$ beyond the
zero contour (dotted).  The resulting cutoff function $\mathfrak{C}$ is indeed
symmetric with the simultaneous change of sign in $n$ and $q$.

\begin{figure}[htb!]
\centering\noindent
\centerline{\includegraphics[width=0.4\columnwidth]{cutintb3.eps}}
\caption{ZP force cutoff function ${\mathfrak{C}}$ in wavenumber space
$(\hat{n}=n\lambda_{\rm{x}}/\lambda,\hat{q}=q\lambda_{\rm{x}}/\lambda)$ ranging
from 1 at its maximum $n=q=0$ to slightly negative outside the zero contour
(dotted) with the contour interval 0.1.}
\label{f:int}
\end{figure}

The Fourier modes $\overbar{\{\ln\reps\}}_n$ for a given repeating waveform are
independent of the wavelength $\lambda$, and so not changed with its rescaling.
As long as the significant scales of variation in $\reps[z]$ are large compared
to the cutoff light wavelength $\lambda/n\gg\lambda_{\rm{x}}$, the integral can
be treated as a constant $\mathfrak{C}=1$, giving a ZP force that decreases
universally as the inverse square of the wavelength $1/\lambda^2$, from the
external factor in Eq.\ (\ref{e:lifx05}).  This ZP force property is validated
in the numerical calculations for permittivity waves described in Section
\ref{s:gradnum}.

Even nonperiodic functional forms might be treated similarly approximated by
discrete Fourier series', like multiple discrete interacting shaped structures
of varying separation.  However an effectively changing Fourier profile with
separation distance for structures of fixed cross section introduces some scale
dependencies into the ZP force modes. Consideration of more general functions is
outside the scope of the present analysis, which is mainly concerned with the
properties of permittivity waves.

A high frequency cutoff is often used for divergent quantum integrals at
extinction scales natural to the problem. Schwinger, DeRaad, and Milton
\citeyearp{Schwinger+DeRaad+Milton1978} introduce an atomic-scale cutoff in
normally divergent integrals that arise in their formulation of the Casimir
effect for interacting atoms, and reproduce reasonably well measured values of
the surface tension and latent heat of vaporization in liquid He at 0K.  Milonni
and Lerner \citeyearp{Milonni+Lerner1992} find that an atomic-scale cutoff is
justified in microscopic atomic theories applying the Ewald-Oseen extinction
theorem for interacting atoms.

Permittivity variations in acoustic waves are due to changes in electronic
vibration or conductive electronic states, as elaborated in Appendix
\ref{a:stacknum}, so a Compton-frequency cutoff seems appropriate for the
problem $\omega_{\rm{x}}=\omega_{\rm{C}}=m_{\rm{e}}c^2/\hslash=7.7634$E20 rad/s,
and is used for the calculations described in this paper. A Compton-frequency
cutoff has been successfully applied with other ZP effects in electronic states,
historically and most famously with the Lamb shift by Hans Bethe
\paren{\citealt{Bethe1947}\semi\citealt[Section 9.6.2]{Mandl+Shaw2010}}. The
same cutoff is found using a quantum regularization procedure in a medium with a
smoothly varying permittivity \citep{Podgornik+Parsegian2004a}.  QED
calculations for electronic processes exhibit a small but finite correlation
length corresponding to the Compton-scattering scale, which can be taken to
arise due to the limited electron recoil from interactions with virtual photons
by the small but finite electron inertia (\citealt[Chapters
8-9]{Peskin+Schroeder1995}; see also \citealt{Cavalleri+Spavieri1989}).

Since permittivity variations in acoustic waves arise with electronic states,
they should flatten out on a scale comparable to that of Compton electron-photon
scattering since the material substance is actually represented by spatially
discrete reradiating electronic oscillators.  Where the permittivity is constant
over a finite scale $\ell$, reflection coefficients become exponentially
attenuated in frequency with their product going as $\exp[-2\omega\ell/c]$ as in
Eq.\ (\ref{e:RRz0}) for a wavenumber $\kappa=\omega/c$, relevant for
high-frequency processes.  The ZP force $\omega$ integral Eq.\ (\ref{e:lifx03})
is governed by the weighting factor $\omega^2\exp[-2\omega\ell/c]$ near the high
frequency cutoff, which peaks at $\omega=c/\ell$ and drops off exponentially
beyond. With a scale for flattening somewhat less than the Compton scattering
wavelength $\ell=\lambda_{\rm{C}}/(2\pi)$, where
$\lambda_{\rm{C}}=h/(m_{\rm{e}}c)=2.4263$E${-}10$ cm, the integral of the
weighting factor $\omega^2\exp[-2\omega\ell/c]$ out to $\omega\rightarrow\infty$
is close to the integral over $\omega^2$ alone sharply cut off at the Compton
frequency $\omega_{\rm{C}}$, or the integral over $\omega^2$ with an imposed
Lorentzian cutoff around $\omega_{\rm{C}}$ out to $\omega\rightarrow\infty$. It
is known that retardation or dispersion can lead to a natural inherent quantum
regularization \citep{Horsley+Philbin2014}.

\section{In a Permittivity Wave}
\label{s:gradnum}

\begin{figure}[htb!]
\centering\noindent
\centerline{\includegraphics[width=0.63\columnwidth]{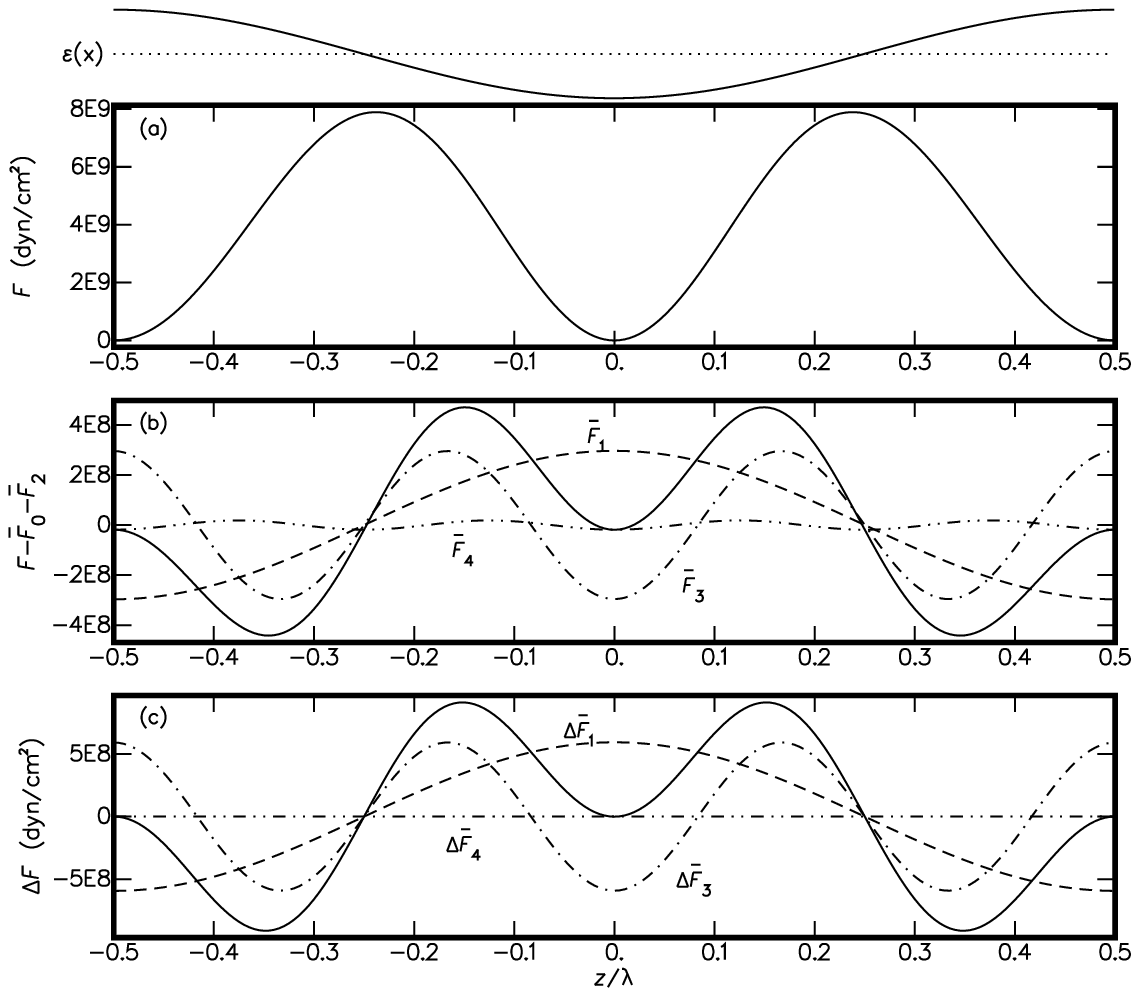}} \caption{ZP
pressure profile $F[z]$ in (a) for a sinusoidal permittivity wave (illustrated
on top), with profiles for the main modes $\bar{F}_0$ and $\bar{F}_2$ subtracted
in (b), and the ZP pressure difference profile $\Delta F[z]$ between the normal
and sign-reversed permittivity waves in (c) (all solid), for wavelength
$\lambda={\rm{E}}{-}6.5$ cm = 3.16 nm and static-permittivity contrast
$\Delta\eps_{\rm{e}}/\langle\eps_{\rm{e}}\rangle=0.1$, and parameters
characteristic of quartz.  Profiles for component modes $\bar{F}_1$,
$\Delta\bar{F}_1$ (dashed), $\bar{F}_3$, $\Delta\bar{F}_3$ (dot-dashed),
$\bar{F}_4$, and $\Delta\bar{F}_4$ (dot-dot-dashed) are also shown.}
\label{f:dem10lm65}
\end{figure}

Figure \ref{f:dem10lm65} shows an example ZP pressure profile $F[z]$ over one
wavelength $\lambda$ in (a) due to a sinusoidal permittivity wave $\reps[z]$ of
representative wavelength and contrast (illustrated on top).  The ZP pressure
profile ranges from near zero to a maximum positive value for expansive in the
most slopping portions of the permittivity wave.

For a permittivity variation symmetric about $z=0$, the Fourier modes
$\overbar{\{\ln\reps\}}_n$ and correspondingly $\bar{F}_n$ are real with
$\bar{F}_{-n}=\bar{F}_n$ in Eq.\ (\ref{e:lifx05}), giving the cosine series for
$n\geq0$ from Eq.\ (\ref{e:Fser})
\begin{equation}
F[z]=\sum_{n=0}{2^{(n!=0)} \bar{F}_n\cos[2\pi nz/\lambda]},
\label{e:Fserc}
\end{equation}
with the logical $(n!=0)$, which is 1 on true and 0 on false, for double weight
on all but the DC mode. Though most of the power in the profile (a) is in the
bulk $\bar{F}_0$ and $\lambda/2$ $\bar{F}_2$ modes, a small residual containing
mainly the $\bar{F}_1$ and $\bar{F}_3$ modes remains, which appears when the
main ZP modes $\bar{F}_0$ and $\bar{F}_2$ are subtacted in (b). The subtraction
is formulated strictly as $F[z]-\bar{F}_0-2\bar{F}_2\cos[4\pi z/\lambda]$, with
$\bar{F}_0$ positive and $\bar{F}_2$ negative.  The ZP force difference between
normal and sign-reversed permittivity waves in (c) appears to be close to two
times the residual in (b).

\begin{figure}[htb!]
\centering\noindent
\centerline{\includegraphics[width=0.63\columnwidth]{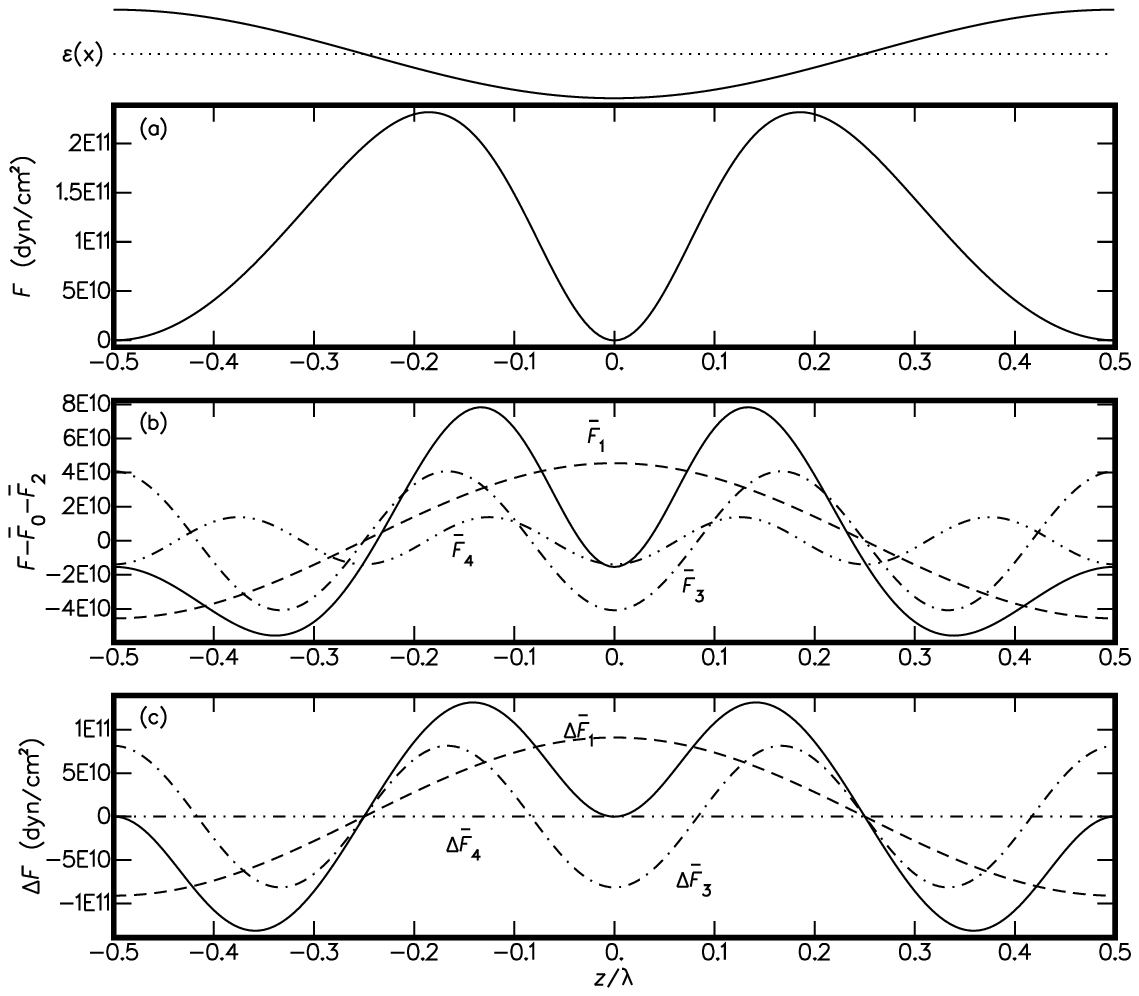}}
\caption{ZP pressure profiles as in Figure \ref{f:dem10lm65} but for a larger
amplitude static-permittivity wave with relative contrast
$\Delta\eps_{\rm{e}}/\langle\eps_{\rm{e}}\rangle=0.5$.}
\label{f:dem03lm65}
\end{figure}

The ZP force model used for the figures is based on the Fourier decomposition
from Section \ref{s:graded}.  The permittivity $\reps[z]$ is defined by Eqs.\
(\ref{e:epsdef}) and (\ref{e:repsdef}) in Appendix \ref{a:stacknum} with a
sinusoidal wave in one parameter as static permittivity
$\eps_{\rm{e}}=\langle\eps_{\rm{e}}\rangle+\Delta\eps_{\rm{e}}\cos2\pi
z/\lambda$, and the sign-reversed wave with $\cos\rightarrow-\cos$.  The
permittivity coefficients $\overbar{\{\ln\reps\}}_n$ in Eq.\ (\ref{e:rhoRpm})
are determined by expanding $\ln[\reps[\cos[2\pi z/\lambda]]]$ as a series in
cosine powers, then substituting with the exponential formula $\cos[2\pi
z/\lambda]=(\exp[i2\pi z/\lambda]+1/\exp[i2\pi z/\lambda])/2$, and collecting
common exponential powers algebraically using Mathematica \citep{Wolfram1991}.
ZP force modes are then calculated using Eq.\ (\ref{e:lifx05}) with the integral
in Matsubara frequency $\xi$ evaluated numerically.  Both Mathematica and C++
were used in different formulations of the problem.

ZP pressure profiles for small amplitude permittivity waves all look
qualitatively like the example shown here for
$\Delta\eps_{\rm{e}}/\langle\eps_{\rm{e}}\rangle=0.1$ .  They are dominated by
the ZP modes $\bar{F}_0$ and $\bar{F}_2$.  ZP pressure profiles for larger
permittivity contrasts do noticeably change in appearance as illustrated in
Figure \ref{f:dem03lm65} for
$\Delta\eps_{\rm{e}}/\langle\eps_{\rm{e}}\rangle=0.5$.  They exhibit an
increased contribution from the ZP wave modes $\bar{F}_1$ and $\bar{F}_3$, with
even a significant contribution from the $\bar{F}_4$ mode.  For both figures,
model parameters representative of quartz are used:\ electronic transition
frequency $\omega_{\rm{e}}=2$E16 rad/s, plasma frequency $\omega_{\rm{p}}=0$,
and mean static permittivity $\langle\eps_{\rm{e}}\rangle=3.5$.

These ZP pressure profiles are in close quantitative agreement with those
obtained by a direct method with the reflection coefficients evaluated by
numerical integration of Eq.\ (\ref{e:Rint}) iterating from a zero prior
approximation, followed by numerical integration of the ZP force in Eq.\
(\ref{e:lifx01}) abruptly truncated at the Compton frequency
$\omega_{\rm{x}}=\omega_{\rm{C}}$. The direct method provides reasonably
indistinguishable results within its noise levels. However direct methods are
more limited as they exhibit a noise cross-talk between Fourier modes close to
the machine precision, which still becomes evident as systematic deviations in
single high $n$ modes or in difference profiles.

Adding previous interations of the reflection coefficient $\tilde{R}$ back into
the integral Eq.\ (\ref{e:Rint}), even for permittivity waves as large as
$\Delta\reps=\langle\reps\rangle$, introduces a change in the mode amplitudes
$\bar{F}_n$ no larger than $2\%$ for the worst cases, at the smallest
wavelengths $\lambda={\rm{E}}{-}9$ cm and in Fourier modes $n\gtrsim30$.
Introducing the denominator in the reflection terms produces relative changes of
E$-6$ in the ZP force in the worst cases, and the perpendicular reflection terms
$\mathcal{R}_s$ produce no larger than an E$-14$ relative effect. Little change
in the profiles is seen at wavelengths near the cutoff, suggesting that the
solution is little affected by the form of the cutoff as with the use of a
Lorentzian cutoff in the Fourier method. In all of the numerical solutions
described in this paper, integrals are evaluated using Simpson's rule with at
least 60 intervals over their ranges of maximum contribution. Increasing the
number of intervals produces no significant changes ${>}1\%$ in the result. Inui
\citeyearp{Inui2008} has applied a direct computational implementation of a
recursive formula for the reflection coefficients to estimate ZP pressures in
linearly graded media.

\begin{figure}[htb!]
\centering\noindent
\centerline{\includegraphics[width=0.63\columnwidth]{fkm03lm65.eps}}
\caption{ZP cosine mode amplitudes $\log|2^{(n!=0)}\bar{F}_n|$ for the ZP
pressure profile in (a), and for the difference profile $\log|2\Delta\bar{F}_n|$
in (b), for waves in the four permittivity parameters:\ static permittivity
$\eps_{\rm{e}}$, electronic transition frequency $\omega_{\rm{e}}$, plasma
frequency $\omega_{\rm{p}}$, and Drude collision frequency $\gamma_{\rm{D}}$.}
\label{f:fk03}
\end{figure}

\begin{figure}[htb!]
\centering\noindent
\centerline{\includegraphics[width=0.63\columnwidth]{ftdem03.eps}}
\caption{ZP cosine-mode amplitudes $\log|2^{(n!=0)}\bar{F}_n|$ in (a) and ZP
pressure differences $\log|2\Delta\bar{F}_n|$ in (b) as functions of wavelength
$\lambda$ for different static-permittivity relative contrasts
$\Delta\eps_{\rm{e}}/\langle\eps_{\rm{e}}\rangle$, showing the main coinciding
modes $|\bar{F}_0|=|2\bar{F}_2|$ (solid), coinciding residuals
$|2\bar{F}_1|=|2\bar{F}_3|$ (dashed), and $|2\bar{F}_4|$ (dot-dot-dashed). To
distinguish overlapping modes, vectors are drawn connecting modes in each
permittivity-contrast group.}
\label{f:ftde}
\end{figure}

In Figure \ref{f:fk03}, mode amplitudes $|2^{(n!=0)}\bar{F}_n|$ are shown for
$\lambda={\rm{E}}{-}6.5$ cm, with a relative contrast of 0.5 in the four
parameters, static permittivity $\eps_{\rm{e}}$, electronic transition frequency
$\omega_{\rm{e}}$, plasma frequency $\omega_{\rm{p}}$, and Drude collision
frequency $\gamma_{\rm{D}}$. Characteristically $|\bar{F}_0|\simeq|2\bar{F}_2|$
and $|2\bar{F}_1|\simeq|2\bar{F}_3|$.  Mode amplitudes drop off with a slope in
$n$ beyond $n=3$ proportional to the relative contrast.

With the reversal in the sign of the permittivity wave, the main Fourier modes
$\bar{F}_0$ and $\bar{F}_2$, with all of the even $n$ modes, do {\it not}
change, and so vanish in the difference profile $\Delta F[z]$.  Both of the next
significant wave modes $\bar{F}_1$ and $\bar{F}_3$, as well as all of the odd
$n$ modes, change sign, and so make up the difference profile $\Delta F[z]$.

Figure \ref{f:ftde} shows ZP pressure mode amplitudes $|2^{(n!=0)}\bar{F}_n|$ as
a function of wavelength $\lambda$ in (a), and ZP pressure differences
$|2\Delta\bar{F}_n|$ in (b), for a number of relative static-permittivity
contrasts $\Delta\eps_{\rm{e}}/\langle\eps_{\rm{e}}\rangle$.  The profiles run
parallel with a slope $1/\lambda^2$, although there is some evidence for
flattening in the slope in the odd modes for $\lambda\lesssim{\rm{E}}{-}8.5$ cm
approaching the Compton cutoff wavelength $\lambda_{\rm{C}}={\rm{E}}{-}9.6$ cm.
Flattening at small wavelengths in both odd and even modes is more evident in
waves in other of the permittivity parameters (not shown). Separation between
contrast groups $\Delta\eps_{\rm{e}}<\langle\eps_{\rm{e}}\rangle$ in the main
Fourier modes $\bar{F}_0$ and $\bar{F}_2$ (solid) go as the relative
permittivity contrast squared, $\bar{F}_1$ and $\bar{F}_3$ (dashed) as the
contrast cubed, and $\bar{F}_4$ (dot-dashed) as the contrast to the fourth
power.

The difference spectra in (b) show only the coinciding residuals $\bar{F}_1$ and
$\bar{F}_3$ (dashed), which are separated by three orders of permittivity
contrast between groups, indicating that the differences, like the modes
themselves, go as the relative permittivity contrast cubed. Only the coinciding
odd modes $\Delta\bar{F}_1$ and $\Delta\bar{F}_3$ appear in the difference
spectra, since ZP modes for $n>4$ are not shown.

Graded solutions for permittivity waves are overall repulsive with the ZP bulk
mode $\bar{F}_0$ always positive, as the product in Eqs.\ (\ref{e:lifx04}) and
(\ref{e:lifx05}) is negative,
$(n-q)q\overbar{\{\ln\reps\}}_{n-q}\overbar{\{\ln\reps\}}_q=-q^2|\overbar{\{\ln\reps\}}_q|^2\leq0$
for $n=0$ and for all $q$, using the Hermitian property
$\overbar{\{\ln\reps\}}_{-q}=\overbar{\{\ln\reps\}}_{q}^*$.  For a permittivity
wave $\rho_{\sop}[z]\sim\cos[2\pi z/\lambda]$ of relatively small amplitude, so
that $\tilde{R}_{\sop\pm}[z]$ can be ignored in the solution Eq.\ (\ref{e:Rint})
of the differential Eq.\ (\ref{e:dRsp}), smoothly varying reflection
coefficients are antisymmetric $R_{\sop+}[z]=-R_{\sop-}[z]$ at relatively high
wavenumbers $\kappa\gg\pi/\lambda$ or high frequencies $\omega/c\gg\pi/\lambda$,
which leads to a repulsive ZP force. Repulsive solutions are similarly found
with soft boundaries modeled by permittivity gradients
\citep{Podgornik+Parsegian2004a, Philbin+Xiong+Leonhardt2010}.

ZP modes above some $n$, as $n>1$ in the examples of Figures \ref{f:dem10lm65}
and \ref{f:dem03lm65}, dip negatively at the lows and highs in the permittivity,
which adds an attractive component that may offset the repulsive force in more
picket-fence-like periodic permittivity waveforms. In the examples described in
this section, the ZP mode $\bar{F}_1$ is anticorrelated with the permittivity
wave and attractive at the wave bottom.  It may act to either pump-up or
dissipate a longitudinal acoustic wave depending upon the sign of the
acousto-optic effect for the material.

\section{Piezo-ZP Energy Dynamics}
\label{s:piezo}

The mechanical energy transfer in a bulk piezoelectric medium associated with
acoustic waves is determined by a total energy $U$ in a stressed piezo crystal
with the energy increment \citep{Mason1966}
\begin{equation}
dU=T_{\imath\jmath}dS_{\imath\jmath}+E_\imath dD_\imath,
\label{e:dU}
\end{equation}
summing over repeated indices in a term, with the usual stress
$T_{\imath\jmath}$ and strain $S_{\imath\jmath}$ symmetric $3\times3$ tensors
containing both longitudinal ($\imath=\jmath$) and transverse
($\imath\not=\jmath$) effects, and the electric field $E_\imath$ and electric
displacement $D_\imath$ 3-element spatial vectors.  Indices count coordinates
$\imath=x,y,z$; $x_\imath$ denotes the spatial coordinate
$(x_x,x_y,x_z)\equiv(x,y,z)$.

For a developed example of the phenomenology of ZP energy coupling, the present
work is limited to nonconducting piezo materials in a bulk medium with no
changes of state over their ranges of variation.  The total energy $U$ from
(\ref{e:dU}) is then conserved, $dU=0$, with the ZP forced mechanical work
$T_{\imath\jmath}dS_{\imath\jmath}$ balancing the capacitive energy loss
$-E_{\imath}dD_{\imath}$.

Material parameters are assumed to be linearly coupled supposing small
variations, and thus connected by two constitutive relations, with the most
convenient pair for this work written \citep[Table 2.1]{Ikeda1990}
\begin{equation}\begin{split}
T_{\imath\jmath}=&c_{\imath\jmath pq} S_{pq}- h_{p\imath\jmath} D_p,\\
E_{\imath}=&-h_{\imath\jmath p} S_{\jmath p}+\backepsilon_{\imath\jmath} D_{\jmath},
\label{e:pz}
\end{split}\end{equation}
with the elastic stiffness at constant electric displacement $c_{\imath\jmath
pq}$ a $3\times3\times3\times3$ tensor, which is defined using the inverse of
the material compliance tensor $s_{\imath\jmath pq}$, with the impermeability at
constant strain $\backepsilon_{\imath\jmath}$ a $3\times3$ matrix, and with the
piezoelectric constant $h_{\imath\jmath p}$ a $3\times3\times3$ tensor, which
enters symmetrically between the material response for the converse piezo effect
in the first equation in (\ref{e:pz}) and the voltage response for the normal
piezo effect in the second equation.

Stress-caused energetic processes do not give a capacitive transfer when the ZP
mechanical work $T_{\imath\jmath}dS_{\imath\jmath}$ done over a cycle vanishes,
with the stress $T_{\imath\jmath}$ and strain $S_{\imath\jmath}$ varying in
phase.  Then the electric field $E_{\imath}$ and electric displacement
$D_{\imath}$ both must also vary in phase for $dU=0$ in Eq.\ (\ref{e:dU}), and
in phase with the stress and strain, as is evident from the constitutive
relations (\ref{e:pz}).  A transfer of energy between the ZP mechanical and
dielectric terms occurs if integrals of $T_{\imath\jmath}dS_{\imath\jmath}$ or
$E_{\imath}dD_{\imath}$ over a cycle are nonzero.  If the stress leads the
strain, the integral of $TdS$ is positive, and ZP mechanical work is converted
into dielectric energy.  The stress may lag the strain too, corresponding to the
transfer from an external electrical supply into ZP opposed mechanical motions.

The ZP pressure $F[z,t]$ in a permittivity wave is described as a longitudinal
$z$ traction force of expansion acting inside the material, and so is
represented in the force-balance equation \citep[Sections 2.B and 2.C]{Auld1973}
\begin{equation}
\rho\frac{\partial^2 u_{\imath}}{\partial t^2}-\frac{\partial T_{\imath\jmath}}{\partial x_{\jmath}}=
-\frac{\partial F[z,t]}{\partial z}\delta_{\imath z},
\label{e:feq01}
\end{equation}
with $\rho$ the density of the medium, and $u_{\imath}$ the
particle-displacement 3 vector.

Particle displacements define the strain tensor
\begin{equation}
S_{\imath\jmath}=\frac{1}{2}\left(\frac{\partial u_{\imath}}{\partial x_{\jmath}}+\frac{\partial u_{\jmath}}{\partial x_{\imath}}\right).
\label{e:Sdef}
\end{equation}
For consistency with the forcing, 1D solutions are studied with
$\partial/\partial x=\partial/\partial y=0$ everywhere in the interior of the
crystal, which gives ZP force driving equations for the three nonvanishing
nonredundant strain elements $S_{xz}$, $S_{yz}$, and $S_{zz}$; as
$S_{\imath\jmath}$ is symmetric, $S_{zx}=S_{xz}$ and $S_{zy}=S_{yz}$.  Taking
the $z$ derivative of (\ref{e:feq01}) yields
\begin{equation}
\rho\frac{\partial^2 S_{zz}}{\partial t^2}-\frac{\partial^2 T_{zz}}{\partial z^2}=
-\frac{\partial^2 F[z,t]}{\partial z^2},
\label{e:feq02a}
\end{equation}
and for $\imath=x,y$
\begin{equation}
2\rho\frac{\partial^2 S_{\imath z}}{\partial t^2}=\frac{\partial^2 T_{\imath z}}{\partial z^2}.
\label{e:feq02b}
\end{equation}

Using the first constitutive relation in (\ref{e:pz}) to eliminate the tension
terms in the force equations (\ref{e:feq02a}) and (\ref{e:feq02b}) yields the
strain for a given ZP force $F$ and electric displacement $D_{\imath}$; for
$S_{zz}$
\begin{equation}
\rho\frac{\partial^2 S_{zz}}{\partial t^2}-(c_{zz\jmath z}+c_{zzz\jmath})\frac{\partial^2 S_{\jmath z}}{\partial z^2}=
\frac{\partial^2}{\partial z^2}\left(-F[z,t]-h_{\jmath zz} D_{\jmath}\right),
\label{e:feq03a}
\end{equation}
and for $S_{xz}$ and $S_{yz}$
\begin{equation}
2\rho\frac{\partial^2 S_{\imath z}}{\partial t^2}
-(c_{\imath z\jmath z}+c_{\imath zz\jmath})\frac{\partial^2 S_{\jmath z}}{\partial z^2}=
-h_{\jmath\imath z}\frac{\partial^2 D_{\jmath}}{\partial z^2}.
\label{e:feq03b}
\end{equation}

The three coupled linear differential equations (\ref{e:feq03a}) and
(\ref{e:feq03b}) are wave equations for the forcing $F$ in three nonzero
independent strain and three electric-displacement elements.  The first equation
is in the direction of the vector forcing, and leads to a particular solution
having the generic form
\begin{equation}
\bar{S}[k_z]=\frac{k_z^2}{k_z^2-k_S^2}\left(s\bar{F}[k_z]+sh\bar{D}[k_z]\right),
\label{e:feq04}
\end{equation}
for the scalar strain $S$ and electric displacement $D$, which are tensor and
vector amplitudes in specific directions appropriate to the overall solution.
The wave equation is written in Fourier space taking $\partial^2/\partial
t^2\rightarrow -4\pi^2\nu^2$, and $\partial^2/\partial
z^2\rightarrow -4\pi^2k_z^2$ with $\bar{S}[k_z]$, $\bar{F}[k_z]$ and
$\bar{D}[k_z]$ functions of $k_z$, the $z$ element of the wavenumber in cycles
per cm, which is related to the dimensionless integer wavenumber $n$ of Section
\ref{s:graded} as $k_z=n/\lambda$.  An effective constant scalar converse piezo
coefficient is denoted $h$, and constant scalar material compliance $s$, which
is the relevant component of the inverse of the elastic stiffness at constant
electric displacement $c_{\imath\jmath pq}$.  The characteristic wavenumber is
defined $k_S=\nu(s\rho)^{1/2}$, for $1/(s\rho)^{1/2}$ the wave phase speed at
vibration frequency $\nu$ with $\rho$ the material density; $k_S$ represents the
vibration wavenumber for the frequency $\nu$ of the ZP forcing.

Strictly, piezo coupling may occur at resonance with $k_z=k_S$ for a vanishing
denominator in Eq.\ (\ref{e:feq04}), but acoustic waves with wavelength smaller
than the dimensions of the crystal will produce a piezo fluctuating voltage with
zero sum effect. Energy transfer with small piezos, comparable in size to the
acoustic wavelength, has been demonstrated experimentally \citep{WangX+++2007},
but for macroscopic piezos, only a bulk modulation produces a single-signed
voltage effect, which even gives a summed total voltage with incoherent
contributions from independent waves through the piezo volume.

In the case where $k_S=0$ as for the ZP bulk forcing, the multiplier in Eq.\
(\ref{e:feq04}) becomes $1$ over the entire Fourier domain (including at
$k_z=0$ by L'Hospital's rule), and transforming back to the spatial domain gives
a strain response $S=sF+shD$.  The ZP bulk force contributes to the strain
directly in the term $sF$, and indirectly with the added converse piezo
contribution $shD$, which is both relatively small and phase shifted, being of
relative amplitude the piezo coupling coefficient \citep[Section
2.6]{Ikeda1990}. The bulk strain solution corresponds to the static response
$T=F$, referring to the constitutive relation (\ref{e:pz}).  A static piezo
response is typically assumed for slow piezo processes usual in
energy-harvesting applications.

A spatially periodic permittivity wave of arbitrary shape of wavelength
$\lambda$ and frequency $\nu$ traveling in $\pm z$ with speed $\lambda\nu$ is
represented by the translation $z/\lambda\rightarrow z/\lambda\mp\nu t$, in each
term in the Fourier series in (\ref{e:lepsser}).  The translation passes through
into the ZP force in the series' Eqs.\ (\ref{e:Fser}) and (\ref{e:Fserc}),
without changing the Fourier modes $\bar{F}_n$ in Eq.\ (\ref{e:lifx04}). As
simply expected, the ZP force travels at velocity $\pm\lambda\nu$ with the
permittivity wave. Just as the $\bar{F}_0$ bulk mode in the unmoving wave is
spatially constant, in the traveling wave it is also temporally constant, and
does not produce a net piezo energy transfer.

A standing permittivity wave is obtained by the superposition of equal-amplitude
oppositely directed $\pm z$ traveling waves, represented by the translation
\begin{equation}
e^{i2\pi n\frac{z}{\lambda}}\rightarrow
\sfrac{1}{2}\left(e^{i2\pi n\left(\frac{z}{\lambda}-\nu t\right)}+e^{i2\pi n\left(\frac{z}{\lambda}+\nu t\right)}\right)=
\sfrac{1}{2}e^{i2\pi n\frac{z}{\lambda}}\left(e^{-i2\pi n\nu t}+e^{i2\pi n\nu t}\right)=
e^{i2\pi n\frac{z}{\lambda}}\cos[2\pi n\nu t],
\nonumber
\end{equation}
in each term of the Fourier series (\ref{e:lepsser}).  In the superposition a
cosine factor containing the time dependence is adjoined to each mode. A purely
sinusoidal wave in the log of the permittivity is defined by just the two
nonzero $n=\pm1$ Fourier modes, which gives a Fourier series for the reflection
coefficient also with only the two nonzero modes $R_{(p\pm)n}$ with $n=\pm1$ in
Eq.\ (\ref{e:rhoRpm}), but a ZP force with three nonzero modes $\bar{F}_0$ and
$\bar{F}_{\pm2}$ from the products between pairs of Fourier terms in Eq.\
(\ref{e:lifx04}).  The temporal factors multiply in the ZP force modes yielding
$\cos[2\pi n\nu t]\cos[2\pi n'\nu t]=\cos[2\pi\nu t]^2=(1+\cos[4\pi\nu t])/2$ in
all the product combinations between mode pairs $n=\pm1$ and $n'=\pm1$,
irrespective of their signs.  A uniform temporal modulation of the three ZP
modes $\bar{F}_0$ and $\bar{F}_{\pm2}$ from 1 to 0 to 1, or from on to off to
on, at the frequency $2\nu$ is introduced, as was suggested in Figure
\ref{f:timeline} in the introduction.  In a nonsinusoidal wave in the log of the
permittivity with more than just the $n=\pm1$ Fourier terms, all of the ZP
modes, including the bulk $\bar{F}_0$, become modulated with differing temporal
factors of more complicated form.

\begin{figure}[htb!]
\centering\noindent
\centerline{\includegraphics[width=0.63\columnwidth]{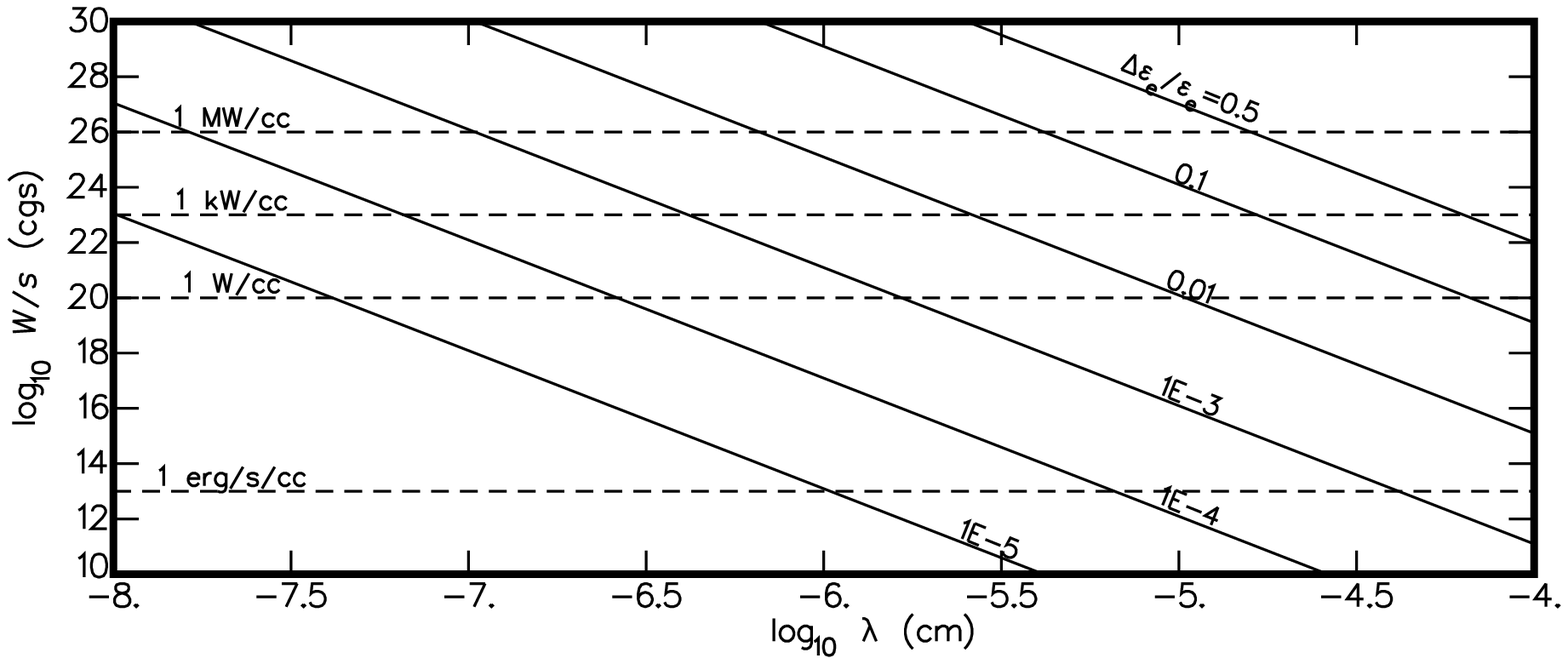}}
\caption{Available ZP power $W$ per unit of crystal volume as a function of
wavelength $\lambda$ in a standing permittivity wave of relative contrast
$\Delta\eps_{\rm{e}}/\langle\eps_{\rm{e}}\rangle$ with quartz acoustic frequency
$\nu$ (on top).  Power levels are shown for quartz (dotted).}
\label{f:wgfde}
\end{figure}

In acting in an elastic body of total length $L$, a ZP bulk pressure $\bar{F}_0$
produces a dimensionless strain $s\bar{F}_0$, which increases the body length by
$s\bar{F}_0L$, where $s$ is a representative elastic compliance for the
material. The ZP force does net work through the length of the elastic body $L$
enlarging it by a continuous motion.  As the ZP force changes in time $t$ as in
Figure \ref{f:timeline} from $\bar{F}_0[t{=}0.25/\nu]=0$ to its extreme value
$\bar{F}_0[t{=}0.5/\nu]=\bar{F}_0$ (not necessarily linearly), it does a total
work defined as the integral of $\bar{F}_0[t]$ times the infinitesimal expansion
distance $sL\partial\bar{F}_0[t]$ over the ZP force range from 0 to its maximum
$\bar{F}_0$, which gives the total work done per unit of surface area
$sL\bar{F}_0^2/2$ in terms of the maximum force $\bar{F}_0$. Thus the total
piezo power transfer or energy per unit of time and volume over the two
permittivity half-wave cycles in each full acoustic wave cycle becomes
\begin{equation}
W=\lambda_{\rm{max}}s\nu\bar{F}_0^2.
\label{e:W01}
\end{equation}
including the piezo energy-transmission coefficient $\lambda_{\rm{max}}<1$,
which is needed as the maximum realizable stress-strain phase shift possible
limits the maximum power that can actually be transferred in a given material
\citep[Sections 2.1 and 2.6.2]{Ikeda1990}.

Power estimates are shown in Figure \ref{f:wgfde} using the ZP force profiles
from Section \ref{s:gradnum}, with the static permittivity $\eps_{\rm{e}}=3.5$
for quartz.  The frequency $\nu=v_{\rm{s}}/\lambda$ scale (shown on top) is for
the quartz sound speed $v_{\rm{s}}=5.8$E5 cm/s.  A representative quartz
material compliance $s=1$E${-}11$ Pa$^{-1}$ = 1E$-$12 cm$^2$/dyn and
energy-transmission coefficient $\lambda_{\rm{max}}=0.002$ are used for the
horizontal power levels. Power levels for piezo conversion are extremely high
for quartz biphonon lattice vibrations $\nu\simeq20$ THz = E13.3 Hz or
$\lambda={\rm{E}}{-}7.6$ cm, far in excess of megawatts per cm$^3$, even for a
relative contrast $\Delta\eps_{\rm{e}}/\langle\eps_{\rm{e}}\rangle=1$E${-}3$,
which is safely below crystal damage levels.  At that frequency, power levels
reach a mass/energy equivalent in less than 0.1 seconds for even an amplitude of
oscillation $\Delta\reps\lesssim0.1\langle\reps\rangle$ !

The voltage difference between electrodes is defined by the electric field, the
order of which is determined by the piezo relation $E\simeq gT\simeq
g\bar{F}_0$, for $g=sh$ the piezoelectric voltage constant; for quartz $g\simeq
7$E${-}5$ Volt cm/dyn.  The voltage difference $E$ increases in proportion to
$\bar{F}_0$, whereas the maximum power goes as $\bar{F}_0^2$, so the maximum
current also goes as $\bar{F}_0$. For a relative static-permittivity contrast
$\Delta\eps_{\rm{e}}/\langle\eps_{\rm{e}}\rangle=1$E${-}3$ with $\nu=20$ THz and
$\lambda={\rm{E}}{-}7.6$ cm in Figure \ref{f:ftde}, the ZP bulk force is
$\bar{F}_0={\rm{E}}8$ dyn/cm$^3$, which produces an electric field $E=7$E3 V/cm,
still somewhat less than the atmospheric breakdown electric field for arcing.

\section{Discussion}
\label{s:disc}

The Casimir effect, the attraction of discrete parallel conducting plates in a
vacuum due to quantum ZP fluctuations characteristically weakens from a
$1/\ell^3$ dependence to a $1/\ell^4$ retarded dependence as the plate
separation increases through $\ell\simeq{\rm{E}}{-}6.2$ cm = 6.3 nm, and weakens
further at still larger separations $\ell\gtrsim{\rm{E}}{-}4$ cm = 1 $\mu$m with
a drop off that steepens with increasing temperature.  The Casimir ZP force in a
stack of discrete plates of alternating permittivity always follows closely the
properties of a single pair of plates, as the ZP force at every location in a
stack is determined by the most nearby plates.

The ZP force in a medium with a {\it smoothly} varying permittivity is found to
exhibit no temperature dependencies.  For a given repeating profile with spatial
scales of variation always larger than the light wavelength for the frequency
cutoff, the ZP force decreases universally as the inverse square $1/\lambda^2$
of the wavelength $\lambda$. In a sinusoidal wave it is repulsive.  In a
static-permittivity wave of wavelength $\lambda={\rm{E}}{-}7.6$ cm = 2.5\AA, it
is about 400 times stronger than the Casimir ZP force in the corresponding
discrete alternating-layer stack of the same repetition scale and
static-permittivity contrast, comparing Figures \ref{f:ftde}a and \ref{f:fe}a.
The ZP force is stronger in a stack of discrete perfectly conducting plates in
alternate layers, but weakens dramatically with more realistic conductivities
comparing with Figure \ref{f:fe}c, and falls off much more rapidly with larger
repetition scale anyway.

The main reason that the ZP force in a medium of smoothly varying permittivity
differs so markedly from that in a stack of discrete plates is that with a
smoothly varying permittivity the ZP force is most affected by the {\it highest}
frequency ZP fluctuations, whereas with finite permittivity steps it is
determined by the {\it lowest} frequency ZP fluctuations.  ZP fluctuations are
exponentially attenuated across a constant permittivity layer, which leads to a
low-frequency cutoff in finite permittivity steps.  On the other hand, the main
ZP force contribution in a graded permittivity comes from near the
high-frequency cutoff for a divergent quantum integral, which is taken to be the
Compton electron-photon scattering frequency, as permittivity effects arise with
electronic states.  A very high light-frequency cutoff corresponding to the
scale of Compton scattering seems physically justified, since a graded
permittivity should be smoothly changing only up to the photo-electric
interaction scale, as the material substance is actually described by spatially
discrete electronic oscillators.

The ZP force in a permittivity wave ranges from near zero at the wave extrema to
a maximum repulsive force in the most sloping portions of the wave, represented
by two most significant equal-amplitude harmonics of bulk repulsive dc
$\bar{F}_0$ and double wavenumber $\bar{F}_2$.  With larger amplitudes of
oscillation, additional harmonics may become significant too, including the mode
$\bar{F}_1$, which matches the acoustic wave itself in wavenumber and phase.  In
a medium with a correct sign of acousto-optic effect, the $\bar{F}_1$ mode may
be able to pump up and sustain a longitudinal acoustic wave or lattice vibration
if it is excited above some minimal amplitude.

In a traveling acoustic wave, the ZP bulk force $\bar{F}_0$ remains constant,
but in a standing wave it is modulated at twice the acoustic wave frequency. If
an acoustic wave is standing and coherent in finite domains through the volume
of a piezo crystal as with biphonon lattice vibrations, the bulk mode may be
significantly modulated and couple piezoelectrically with an external electrical
dc load through end electrodes perpendicular to the wave direction, as in the
illustrated `crystal power' layout of Figure \ref{f:piezo}.  A transfer of
electrical energy out of the ZP fluctuating field is predicted with the
possibility of extremely high power levels as summarized in Figure
\ref{f:wgfde}.  In an acoustic wave of frequency 20 THz, in the range of
biphonon lattice vibrations in quartz, estimated power levels reach a mass
energy equivalent $mc^2$ in a crystal mass $m$ of stimulated domains in less
than 0.1 seconds with a relative permittivity contrast still ${\lesssim}0.1$ or
a relative index-of-refraction contrast ${\lesssim}0.3$.

\section*{Acknowledgements}\noindent
It is the author's contention that specific clues for this work have been
Divinely given. The author wishes that the guidance, concepts, and designs for
the extraction of energy from the ZP field described here remain unpatentable,
but stay in the public domain.

\section*{Appendices}
\appendix
\section{Alternating-Layer Stack}
\label{a:stacknum}

\begin{figure}[htb!]
\centering\noindent
\centerline{\includegraphics[width=0.63\columnwidth]{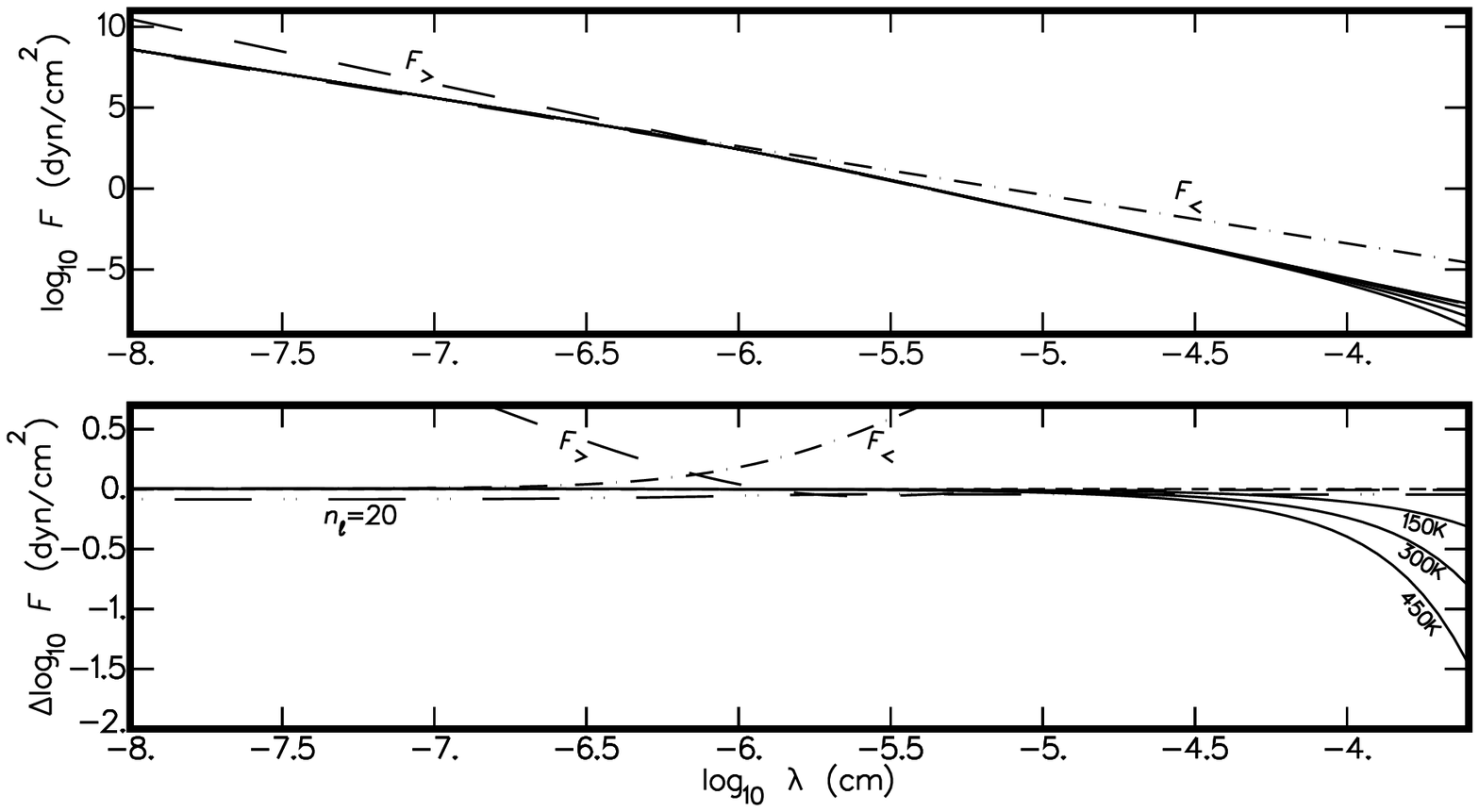}}
\caption{ZP pressure as a function of repetition length $\lambda=2\ell$, from
1E$-$8 cm = 1\AA\ to E$-$3.6 cm = 2.5$\mu$m, in an alternating static
permittivity of relative contrast
$\Delta\eps_{\rm{e}}/\langle\eps_{\rm{e}}\rangle=0.01$, for 3 layers for
$\Theta=0$K and three $\Theta>0$K temperatures (solid lines), and for 41 layers
for 0K ($n_{\ell}=20$, long-dashed), compared with asymptotic power laws $F_-$
and $F_+$ (dashed). ZP pressure with $\log F$ for 3 layers at 0K subtracted is
shown in (b) on a magnified ZP pressure scale.}
\label{f:ftemp}
\end{figure}

Figure \ref{f:ftemp} illustrates ZP pressure as a function of spatial repetition
scale $\lambda=2\ell$ in an alternating-layer stack with equal layer
thicknesses.  The $\Theta=0$K calculations are derived by integrating
numerically in Eq.\ (\ref{e:lifx01}), and the three-layer calculations for three
nonzero temperatures by evaluating the Matsubara sum from (\ref{e:lif}); the
Lifshitz asymptotic power-law line $F_-$ is taken from (\ref{e:Flt02}) and $F_+$
from (\ref{e:Fgt01}) in Appendix \ref{a:limits}.  The three-layer (solid line)
closely follows the relatively small-scale $F_-$ and large-scale $F_+$
asymptotic power laws.  The multilayer 0K solution ($n_{\ell}=20$, long-dashed)
deviates slightly.  Temperature effects become important for repetition scales
$\lambda\gtrsim{\rm{E}}{-}4$ cm = 1$\mu$m.

For all the numerical models described in this paper, a usual form for the
complex permittivity as a function of light frequency $\xi$ is adopted
\begin{equation}
\eps[\xi]=1+\frac{\eps_{\rm{e}}-1}{1-(\xi/\omega_{\rm{e}})^2} -\frac{\omega_{\rm{p}}^2}{(\xi+i\xi_0)(\xi+i\gamma_{\rm{D}})},
\label{e:epsdef}
\end{equation}
which represents a single electronic band at frequency $\omega_{\rm{e}}$ of
strength $\eps_{\rm{e}}-1$, and an added conductive effect for the plasma
frequency $\omega_{\rm{p}}$, with the Drude collision frequency
$\gamma_{\rm{D}}$.  A small base frequency $\xi_0$ is included for numerical
calculations.

The static permittivity $\eps_{\rm{e}}$ is the permittivity at zero frequency
$\xi=0$ in a nonconductor $\omega_{\rm{p}}=0$; $\eps_{\rm{e}}=3.5$ for quartz.
The permittivity at the high frequency extreme exhibits the limit
$\eps[\xi{\rightarrow}\infty]=1$, which is found to be applicable even in
generalized conductivity models that include contributions from free electrons
\citep{Klimchitskaya+++2009}. Characteristically the highest electronic
transition is most important for ZP forces $\omega_{\rm{e}}\simeq 2$E16 rad/s,
corresponding to the Rydberg frequency
$\omega_{\rm{e}}=m_{\rm{e}}e^4/\hslash^3/2$, for $m_{\rm{e}}$ the electron rest
mass and $e$ the elementary charge.  A lattice vibration band at around 9E12
rad/s in crystals is the next lower-frequency effect, but is of relatively small
oscillator strength to make a significant contribution. The plasma frequency
$\omega_{\rm{p}}$ ranges from 0 to 2E15 rad/s in semiconductors, and reaches
about 1.6E16 rad/s in doped semiconductors or in the best room-temperature
metallic conductors; in normal thick conductors $\gamma_{\rm{D}}\simeq2$E14
rad/s.  In numerical calculations with a nonzero plasma frequency
$\omega_{\rm{p}}$, integrals are started a little above $\xi=0$ or a small base
frequency $\xi_0\lesssim{\rm{E}}3$ rad/s is introduced to avoid infinities at
$\xi=0$, which are procedures that only insensitively affect the results.

Due to the Kramers-Kr{\"o}nig causality constraints, the complex permittivity is
real on its imaginary frequency axis consistent with the projected
$\reps[\xi]\equiv\eps[i\xi]$ from Eq.\ (\ref{e:epsdef})
\begin{equation}
\reps[\xi]=1+\frac{\eps_{\rm{e}}-1}{1+(\xi/\omega_{\rm{e}})^2}
+\frac{\omega_{\rm{p}}^2}{(\xi+\xi_0)(\xi+\gamma_{\rm{D}})}.
\label{e:repsdef}
\end{equation}
The projected real permittivity is a slowly decreasing monotonic function of
frequency $\xi$; it is plotted in many places for conductors or nonconductors
(e.g.\ \citealt[Figure 20]{Klimchitskaya+++2009}).

The parametric representation in (\ref{e:epsdef}) and (\ref{e:repsdef}) is taken
to remain applicable over the range of permittivity change.  Thus small
permittivity changes $|\Delta\reps|$ around the average $\langle\reps\rangle$,
$|\Delta\reps|\ll\langle\reps\rangle$, can be represented by the derivative
expansion
\begin{equation}
\Delta\reps=\frac{\partial\reps}{\partial\eps_{\rm{e}}}\Delta\eps_{\rm{e}}
+\frac{\partial\reps}{\partial\omega_{\rm{e}}}\Delta\omega_{\rm{e}}
+\frac{\partial\reps}{\partial\omega_{\rm{p}}}\Delta\omega_{\rm{p}}
+\frac{\partial\reps}{\partial\gamma_{\rm{D}}}\Delta\gamma_{\rm{D}}.
\label{e:Deps}
\end{equation}
Figure \ref{f:ftemp} illustrates the ZP pressure that arises with variations in
the static permittivity between even and odd layers
$\eps_{\rm{e0}}=\langle\eps_{\rm{e}}\rangle-\Delta\eps_{\rm{e}}$ and
$\eps_{\rm{e1}}=\langle\eps_{\rm{e}}\rangle+\Delta\eps_{\rm{e}}$, with
$\langle\eps_{\rm{e}}\rangle=3.5$ and
$\Delta\eps_{\rm{e}}=0.01\langle\eps_{\rm{e}}\rangle$, for
$\omega_{\rm{e}}=2$E16 rad/s in a nonconductor $\omega_{\rm{p}}=0$.

\begin{figure}[htb!]
\centering\noindent
\centerline{\includegraphics[width=0.63\columnwidth,bb=25 137 575 565]{fefwe.eps}}
\centerline{\includegraphics[width=0.63\columnwidth,bb=25 337 575 560.2]{fwpv.eps}}
\caption{ZP pressure $F[\lambda=2\ell]$ with different relative contrasts,
alternating in static permittivity $\eps_{\rm{e}}$ in (a), electronic transition
frequency $\omega_{\rm{e}}$ in (b), and plasma frequency with $\omega_{\rm{p1}}$
in odd layers in (c). Calculations are shown for 41 nearby layers
($n_{\ell}=20$, solid) and the minimum case of 3 layers ($n_{\ell}=1$, dashed).}
\label{f:fe}
\end{figure}

Figure \ref{f:fe} shows the ZP pressure $F[\lambda]$ alternating the static
permittivity in (a) around $\langle\eps_{\rm{e}}\rangle=3.5$, with
$\omega_{\rm{e}}=2$E16 rad/s and $\omega_{\rm{p}}=0$, the electronic transition
frequency in (b) around $\langle\omega_{\rm{e}}\rangle=2$E16 rad/s, with
$\eps_{\rm{e}}=3.5$ and $\omega_{\rm{p}}=0$, and the plasma frequency in (c)
between $\omega_{\rm{p0}}=0$ in nonconducting even layers and $\omega_{\rm{p1}}$
in conducting odd layers, using $\gamma_{\rm{D}}=2$E14 rad/s and
$\eps_{\rm{e}}=1$ supposing the ideal of evacuated nonconducting layers for
illustration. In all of the examples, only quite small changes are seen with
more than a minimal number of layers $n_{\ell}\gtrsim1$.  Shifting the spatial
phase by one layer thickness, that is interchanging the even and odd layers,
produces very small differences, which go only as third order or less in the
permittivity contrast.

Symmetric multilayer stacks all follow the thin limit for the three-layer
sandwich solutions discussed in Appendix \ref{a:limits}, going as $1/\lambda^3$
or $1/\ell^3$.  However the spatially varying electronic transition frequency
$\omega_{\rm{e}}$ in (b) exhibits a different definite thick-layer power law for
retarded solutions going as $1/\ell^8$ rather than the $1/\ell^4$ predicted for
usual circumstances, as discussed in Appendix \ref{a:limits}.  For all relative
static-permittivity contrasts, the ZP force in an alternating layer stack of
repetition scale $\lambda={\rm{E}}{-}7.6$ cm = 2.5\AA\ is only about 1/400 that
found in the corresponding sinusoidal permittivity wave, comparing ZP pressures
between Figures \ref{f:fe}a and \ref{f:ftde}a.

The ZP force increases with relative permittivity contrast, going as
$(\Delta\eps_{\rm{e}}/\langle\eps_{\rm{e}}\rangle)^2$ in (a), and as
$(\Delta\omega_{\rm{e}}/\langle\omega_{\rm{e}}\rangle)^2$ in (b), with
deviations no larger than 0.5\% in relative strength $\Delta F/F$ over the range
of repetition scales up to relative contrasts 0.01.  The ZP force follows a
squared dependence since the frequency integrals are over terms
$R_{\sop+}[{\succ}\sfrac{-\ell}{2}] R_{\sop-}[{\prec}\sfrac{\ell}{2}]$ as in
(\ref{e:RRz0}), where $R_{\sop+}[{\succ}\sfrac{-\ell}{2}]\sim
r_{\sop}[\sfrac{-\ell}{2}]$ and $R_{\sop-}[{\prec}\sfrac{\ell}{2}]\sim
r_{\sop}[\sfrac{\ell}{2}]$ from the innermost jump, with
$r_{\sop}[\pm\sfrac{\ell}{2}]\sim\Delta\reps/\langle\reps\rangle$ from Eq.\
(\ref{e:rsp}) for small permittivity contrasts as in the derivatives for graded
media in (\ref{e:rhosp}).

ZP forces for alternating plasma frequencies $\omega_{\rm{p1}}$ in \ref{f:fe}c
all follow thin asymptotic $1/\ell^3$ power laws up to a scale $\lambda=2\ell$,
where the solution joins the upper $1/\ell^4$ limiting power-law line. Solutions
with higher plasma frequencies $\omega_{\rm{p}}>{\rm{E}}20$ rad/s follow
essentially this limiting solution line, which corresponds closely to the
Casimir-predicted perfect-conductivity limiting form in Eq.\ (\ref{e:Fgt02}).

Using repetition length $\lambda=\ell_0+\ell_1$ to characterize
alternating-layer stacks facilitates comparison with wavelength in a
permittivity wave.  The figures show only alternating-layer calculations with
$\ell_0=\ell_1=\ell=\lambda/2$, as the Casimir ZP force with differing even- and
odd-layer thicknesses is determined essentially by the even (middle-) layer
thickness alone, and exhibits little change with increasing relative thickness
above some minimal $\ell_1\lesssim\ell_0$.  Likewise, the ZP force exhibits only
a weak dependence upon the number of layers $n_{\ell}$ used in the calculation,
with deviations no larger than 0.01 in $\log F$ or 2.3\% in relative strength
over the range of repetition scales ($\Delta F/F=\Delta\log F/\log e$).  The
interfacial-reflection contributions from jumps in the reflection coefficients
$R_{\sop\pm}[z]$ in a multilayer stack are increasingly exponentially attenuated
with increasing distance from the middle $z=0$ layer in Eq.\ (\ref{e:Rsp-}).

\begin{figure}[htb!]
\centering\noindent
\centerline{\includegraphics[width=0.63\columnwidth,bb=25 137 575 579]{wfefwe.eps}}
\centerline{\includegraphics[width=0.63\columnwidth,bb=25 341 575 560.2]{wfwp.eps}}
\caption{Available ZP power $W$ per unit of crystal volume as a function of
repetition length $\lambda=2\ell$ in a stack with alternating static
permittivity in (a), electronic transition frequency in (b), and plasma
frequency in (c), supposing a modulation frequency $\nu=v_{\rm{s}}/\lambda$ with
$v_{\rm{s}}=5.8$E5 cm/s, showing power levels (dotted).}
\label{f:wf}
\end{figure}

Figure \ref{f:wf} shows estimates for the available ZP power in alternating
layer stacks. Nominally a modulation frequency $\nu=v_{\rm{s}}/\lambda$ (shown
on top) with speed $v_{\rm{s}}=5.8$E5 cm/s is adopted representative of quartz,
but power levels scale in proportion to frequency for schemes that modulate
layers at a different frequency.  For an alternating conductivity in (c), levels
are adjusted downward slightly to take account of intermediate quartz piezos in
even layers with $\eps_{\rm{e0}}=3.5$ unmodulated in permittivity.

Power estimates are for cycling the permittivity state at a rate $\nu$ rather
than $2\nu$, and so follow the permittivity-wave formula Eq.\ (\ref{e:W01})
divided by 2.  Also since the plasma frequency is cycled only in odd layers with
a passive quartz piezo in between for (c), only half the volume participates in
the energy extraction, so those power estimates follow the permittivity-wave
formula Eq.\ (\ref{e:W01}) divided by 4.  The material compliance $s=1$E${-}11$
Pa$^{-1}$ = 1E$-$12 cm$^2$/dyn and energy-transmission coefficient
$\lambda_{\rm{max}}=0.002$ are used for the horizontal power levels for quartz
or typical piezo materials.

The available power goes like the square of the ZP force and so drops quite
rapidly with increasing $\lambda$, and also depends quite strongly on the
relative permittivity contrast in (a) and (b) going as the fourth power. The
most favorable power estimates are found with the plasma-frequency modulation in
(c).  For the example described in the introduction, a perfect conductor with
intermediate quartz piezos shows an available power of about 3.5 kW/cm$^3$ for
$\lambda={\rm{E}}{-}6$ cm = 10 nm, or $\ell=5$ nm, modulated at the nominal
frequency $\nu=5.8$E11 Hz.  The higher modulation frequency $\nu={\rm{E}}13=10$
THz gives an available power that scales up to about 60 kW/cm$^3$.  However
power levels fall off very rapidly with less-than-perfect conductivity.  For
more realistic conductivities for semiconductors with
$\omega_{\rm{p}}\lesssim2$E16, the power level in the example drops to ${<}120$
W/cm$^3$.  Layers need to be considerably thinner for effects to become very
appreciable.

\section{Asymptotic Three-Layer Solutions}
\label{a:limits}

The Lifshitz collaboration \citep{Lifshitz1956, Dzyaloshinskii+++1961} reports
asymptotic forms for the ZP force in a three-layer stack for middle layers
$\ell_0$ thinner and thicker than a critical cross-over thickness
$\ell_{\pm}\simeq c/(\omega_{\rm{e}}\sqrt{\eps_{\rm{e0}}})$, for layers still
thinner than the light wavelength for the Matsubara base frequency where
temperature becomes important; $\eps_{\rm{e0}}$ is the static permittivity or
$\sqrt{\eps_{\rm{e0}}}$ the index of refraction in the middle layer.  The
highest frequency for permittivity change is the electronic transition frequency
$\omega_{\rm{e}}$, supposing a lower plasma frequency
$\omega_{\rm{p}}<\omega_{\rm{e}}$ in Eq.\ (\ref{e:repsdef}). Above that
frequency the permittivity becomes asymptotic $\eps\rightarrow1$ and light waves
are unretarded. The cross-over scale $\ell_{\pm}$ is the smallest wavelength for
light travel in the medium for which retardation effects are seen, and thus
determines the minimum layer thickness for the so-called `retarded' solutions.
In this section derivations for the limiting profiles are described and results
discussed in the context of permittivity variations due to different of the
permittivity parameters from Eq.\ (\ref{e:repsdef}).

The ZP force integral (\ref{e:lifx01}) is limited in its frequency range due to
the exponential attenuation factor $\exp[-2\kappa_0\ell_0]$, which appears in
the reflection terms (\ref{e:RRsp}) in the product of reflection coefficients in
a layer $\ell_0$ of constant permittivity in Eq.\ (\ref{e:RRz0}).  When the
middle layer of a three-layer stack is sufficiently thin
$\ell_0<c/(\omega_{\rm{e}}\sqrt{\eps_{\rm{e0}}})$, the corresponding critical
wavenumber from the exponential factor is large
$\kappa_0=1/\ell_0>\omega_{\rm{e}}\sqrt{\eps_{\rm{e0}}}/c$. Such large
wavenumbers change relatively little between layers $\jmath$, as the wavenumber
$\kappa_{\jmath}[\xi,\omega]=(\reps_{\jmath}[\xi]\xi^2+\omega^2)^{1/2}/c$ is
either determined by a frequency $\omega\gtrsim\omega_{\rm{e}}$, which is the
same between layers $j$, or by a Matsubara frequency $\xi\gtrsim\omega_{\rm{e}}$
above the highest frequency for significant permittivity $\reps_{\jmath}[\xi]$
change.  With a spatially constant wavenumber, the interfacial reflections at
the permittivity jumps $r_{s\pm}[\pm\sfrac{\ell_0}{2}]$ from (\ref{e:rsp})
become negligible, and the perpendicular $s$ reflection coefficient
$R_{s\pm}[\pm\sfrac{\ell_0}{2}]$ and reflection term $\mathcal{R}_s$ can be
ignored.  Wavenumbers $\kappa_{\jmath}$ divide out in the parallel $p$
polarization interfacial reflections leaving just the permittivities
$\reps_{\jmath}[\xi]$ in $r_{p\pm}[\pm\sfrac{\ell_0}{2}]$ from (\ref{e:rsp}).

Dropping the product of reflection coefficients from the denominator of
$\mathcal{R}_p$ in (\ref{e:RRsp}), as it is fourth-order in the relative
permittivity contrast compared to the main second-order effect, leaves the only
$\kappa_0$ dependence under the integral in the term $\eta^2\exp[-2\eta]$,
taking $\eta=\kappa_0\ell_0$ to be a substitute variable of integration for
$\omega$ as in Eq.\ (\ref{e:lifx02}).  As the permittivity $\reps_{\jmath}[\xi]$
does not depend upon $\omega$, it does not depend upon its substitute
independent variable $\eta$, and the integrals separate leaving for the $\eta$
integral the Gamma function $\int_0^\infty{\eta^2e^{-2\eta}d\eta}=\sfrac{1}{4}$,
then
\begin{equation}
F_-=-\frac{\hslash}{8\pi^2\ell_0^3}
\int_0^\infty{\left(\frac{\Delta\reps[\xi]}{\langle\reps[\xi]\rangle}\right)^2 d\xi},
\label{e:Flt01}
\end{equation}
which is the Lifshitz thin-layer negative or compressive ZP pressure written for
permittivities symmetric about a middle layer $\jmath=0$ \citep[Eq.\
(4.18)]{Dzyaloshinskii+++1961}.  The permittivity amplitude of variation
$\Delta\reps[\xi]=(\reps_1[\xi]-\reps_0[\xi])/2$, and average
$\langle\reps[\xi]\rangle=(\reps_0[\xi]+\reps_{1}[\xi])/2$, are used for
consistency with Figure \ref{f:stack}, with a symmetric arrangement
$\reps_{1}[\xi]=\reps_{-1}[\xi]$ for layers numbered $0$, $\pm1$.

If the relative permittivity variation arises due to a varying static
permittivity between layers $\eps_{\rm{e}\top{0}{1}}$ in $\reps[\xi]$ in Eq.\
(\ref{e:repsdef}), with the permittivity amplitude of variation
$\Delta\reps=(\partial\reps[\xi]/\partial\eps_{\rm{e}})\Delta\eps_{\rm{e}}=\Delta\eps_{\rm{e}}/(1+(\xi/\omega_{\rm{e}})^2)$,
and average $\langle\reps[\xi]\rangle=
(\reps[\xi;\eps_{\rm{e}}{\rightarrow}\langle\eps_{\rm{e}}\rangle-\Delta\eps_{\rm{e}};\omega_{\rm{p}}{\rightarrow}0]+
\reps[\xi;\eps_{\rm{e}}{\rightarrow}\langle\eps_{\rm{e}}\rangle+\Delta\eps_{\rm{e}};\omega_{\rm{p}}{\rightarrow}0])/2=
1+(\langle\eps_{\rm{e}}\rangle-1)/(1+(\xi/\omega_{\rm{e}})^2)$, the $\xi$
integral in (\ref{e:Flt01}) can be evaluated yielding
\begin{equation}
F_-=-\frac{\hslash\omega_{\rm{e}}\sqrt{\eps_{\rm{e}}}}{32\pi\ell_0^3}
\left(\frac{\Delta\eps_{\rm{e}}}{\langle\eps_{\rm{e}}\rangle}\right)^2,
\label{e:Flt02}
\end{equation}
which is the limiting form shown in Figure \ref{f:ftemp}.

The thin-layer asymptotic ZP pressure Eq.\ (\ref{e:Flt01}) for a varying plasma
frequency $\omega_{\rm{p}}$ with static permittivity $\eps_{\rm{e}}=1$ exhibits
a straightforward solution form too (not written out here).  An unmanageable
analytic integral is arrived at with a varying electronic transition frequency
$\omega_{\rm{e}}$ in a nonconductor $\omega_{\rm{p}}=0$, and a nonanalytic
integral with a varying Drude collision frequency $\Gamma_{\rm{p}}$ with static
permittivity $\eps_{\rm{e}}=1$.  Anyway the numerical examples from Figure
\ref{f:fe} show that all these follow the asymptotic thin-layer power law
$1/\ell_0^3$ for small $\ell_0$, although conductors exhibit a cross-over scale that
varies greatly with plasma frequency $\omega_{\rm{p}}$.

When the middle layer is sufficiently thick
$\ell_0>c/(\omega_{\rm{e}}\sqrt{\eps_{\rm{e0}}})$, the critical wavenumber from
the exponential attenuation factor $\exp[-2\kappa_0\ell_0]$ is small
$\kappa_0=(\reps_0[\xi]\xi^2+\omega^2)^{1/2}/c<\omega_{\rm{e}}\sqrt{\eps_{\rm{e0}}}/c$,
and the integral is determined by low-frequency contributions near
$\xi=\omega=0$, but still above the Matsubara $m=1$ frequency.  If
zero-frequency permittivity variations exist, then it seems a good approximation
to suppose that the permittivity is constant in frequency and well approximated
by its zero-frequency value $\reps_{\jmath}[\xi]\rightarrow\reps_{\jmath}[0]$.
Following the Lifshitz approach, new dimensionless integration variables are
substituted into the integral Eq.\ (\ref{e:lifx01}),
$\hat{\xi}=\xi\ell_0\sqrt{\reps_0[0]}/c$ for the frequency $\xi$, and
$\hat{\omega}=\omega\ell_0/c$ for $\omega$, which gives
$\kappa_0=(\hat{\xi}^2+\hat{\omega}^2)^{1/2}/\ell_0$ for the ZP force in the
middle layer.  With the substitutions, all scale dependencies factor out of the
integrals into an $\ell_0^4$ external divisor.  In a symmetric arrangement, only
a weak dependence upon the zero-frequency relative permittivity contrast between
the layers $\Delta\reps[0]/\langle\reps[0]\rangle$ is left in the jump
reflection coefficients in the remaining integral, which leads to
\begin{equation}
F_+=-\frac{\pi^2\hslash c\ \phi_{\rm{dd}}}{240\sqrt{\reps_0[0]}\ell_0^4}
\left(\frac{\Delta\reps[0]}{\langle\reps[0]\rangle}\right)^2.
\label{e:Fgt01}
\end{equation}
quoting the result from the Lifshitz collaboration \citep[Eq.\
(4.22)]{Dzyaloshinskii+++1961}; $\phi_{\rm{dd}}$ is a weak function of the
permittivity ratio
$\reps_1[0]/\reps_0[0]=(\langle\reps[0]\rangle+\Delta\reps[0])/(\langle\reps[0]\rangle-\Delta\reps[0])$,
$\phi_{\rm{dd}}\simeq0.35$ for $\reps_1[0]/\reps_0[0]<3$. For small permittivity
contrasts $\Delta\reps[0]/\langle\reps[0]\rangle\ll1$ as in the numerical
examples shown in Appendix \ref{a:stacknum}, the permittivity ratio
$\reps_1[0]/\reps_0[0]$ is always close to 1, so $\phi_{\rm{dd}}=0.35$. This ZP
force $F_+$ limiting line is shown in Figure \ref{f:ftemp}.  Like what is seen
in Figure \ref{f:fe}b, the asymptotic thick-layer solution Eq.\ (\ref{e:Fgt01})
does not necessarily apply to permittivity variations produced with a varying
electronic transition frequency $\omega_{\rm{e}}$ between layers, as those
permittivity variations vanish at low frequency $\xi$.

Comparing with Figure \ref{f:fe}c, the asymptotic formula Eq.\ (\ref{e:Fgt01})
can be seen to apply too above some infinitesimal repetition period $\lambda$ or
middle-layer cross-over thickness $\ell_0$ in perfect conductors idealized with
$\omega_{\rm{p}}\rightarrow\infty$, or above a finite cross-over thickness
$\ell_0$ that increases in normal conductors with decreasing $\omega_{\rm{p}}$,
at least up to the Matsubara $m=1$ thickness where temperature effects become
important.  With a divergent zero-frequency permittivity
$\lim_{\xi\rightarrow0}\reps_{\pm1}[\xi]\rightarrow\infty$, but with
$\reps_0[0]=\eps_{\rm{e0}}$, it follows $\reps_1[0]/\reps_0[0]\rightarrow\infty$
in a symmetric stack $\reps_1[0]=\reps_{-1}[0]$, giving $\phi_{\rm{dd}}=1$
\citep[Figure 10]{Dzyaloshinskii+++1961}, and the permittivity contrast
asymptotically becomes
$\Delta\reps[0]/\langle\reps[0]\rangle=(\reps_1[0]/\reps_0[0]-1)/(\reps_1[0]/\reps_0[0]+1)\rightarrow1$,
implying
\begin{equation}
F_+=\frac{\pi^2 \hslash c}{240\sqrt{\eps_{\rm{e0}}}\ell_0^4},
\label{e:Fgt02}
\end{equation}
which contains the original Casimir \citeyearp{Casimir1948} solution for an
evacuated middle layer with $\eps_{\rm{e0}}=1$.

The retarded cross-over thickness between the two power laws $\ell_{\pm}$ is
defined strictly by the relation
$F_-[\ell_0{=}\ell_{\pm}]=F_+[\ell_0{=}\ell_{\pm}]$, which gives
\begin{equation}
\ell_{\pm}=1.448\frac{c}{\omega_{\rm{e}}\eps_{\rm{e}}},
\label{e:llg1}
\end{equation}
substituting with Eqs.\ (\ref{e:Flt02}) and (\ref{e:Fgt01}) for
static-permittivity variations with $\Delta\reps[0]=\Delta\eps_{\rm{e}}$ and
$\langle\reps[0]\rangle=\langle\eps_{\rm{e}}\rangle$.  An extra
index-of-refraction $\sqrt{\eps_{\rm{e}}}$ enters into the divisor in this
integral solution compared to the critical cross-over thickness $\ell_{\pm}$ for
the peak wavenumber under the integral introduced at the beginning of this
section. For $\eps_{\rm{e}}=3.5$ and $\omega_{\rm{e}}=2$E16 rad/s,
$\ell_{\pm}=6.20$ nm or $\lambda=2\ell_{\pm}=12.40$ nm = E$-$5.91 cm, which
agrees with the intersection of the dashed lines seen in Figure \ref{f:ftemp}
and the cross-over thickness between the two power laws in Figure \ref{f:fe}a.


\end{document}